\documentclass[%
 reprint,
 amsmath,amssymb,
 aps,
prb,
nofootinbib,
]{revtex4-2}

\usepackage{tabularx}
\usepackage{hyperref} 
\hypersetup{colorlinks=true, citecolor=blue, urlcolor=blue, linkcolor=blue} \usepackage{siunitx}
\usepackage{graphicx}
\usepackage{bm}

\begin{document}


\title{Machine learning search for stable binary Sn alloys with Na, Ca, Cu, Pd, and Ag}

\author{Aidan Thorn, Daviti Gochitashvili, Saba Kharabadze, and Aleksey N. Kolmogorov}\email{kolmogorov@binghamton.edu}
\affiliation{
  Department of Physics, Applied Physics and Astronomy,\\ Binghamton University,         State University of New York, \\PO Box 6000, Binghamton, New York 13902-6000, USA.
}
%

\begin{abstract} 

    We present our findings of a large-scale screening for new synthesizable materials in five M-Sn binaries, M = Na, Ca, Cu, Pd, and Ag. The focus on these systems was motivated by the known richness of M-Sn properties with potential applications in energy storage, electronics packaging, and superconductivity. For the systematic exploration of the large configuration space, we relied on our recently developed {\small MAISE-NET} framework that constructs accurate neural network interatomic potentials and utilizes them to accelerate {\it ab initio} global structure searches. The scan of over two million candidate phases at a fraction of the typical {\it ab initio} calculation cost has uncovered 29 possible intermetallics thermodynamically stable at different temperatures and pressures (1 bar and 20 GPa). Notable predictions of ambient-pressure materials include a simple hP6-NaSn$_2$ phase, fcc-based Pd-rich alloys, tI36-PdSn$_2$ with a new prototype, and several high-temperature Sn-rich ground states in the Na-Sn, Cu-Sn, and Ag-Sn systems. Our modeling work also involved {\it ab initio} (re)examination of previously observed M-Sn compounds that helped explain the entropy-driven stabilization of known Cu-Sn phases. The study demonstrates the benefits of guiding structure searches with machine learning potentials and significantly expands the number of predicted thermodynamically stable crystalline intermetallics achieved with this strategy so far.

\end{abstract}

\maketitle

\section{Introduction}

\textit{Ab initio} screening of vast chemical spaces has become an integral part of materials discovery. The AFLOW~\cite{AFLOW}, Materials Project~\cite{MatProj}, OQMD~\cite{OQMD}, and other open repositories contain \textit{ab initio} results for hundreds of thousands of compounds in observed structure types and demonstrate that density functional theory (DFT) approximations offer a reliable determination of materials’ stability. Mining these databases for new synthesizable compounds or materials with targeted properties has led to numerous interesting predictions~\cite{Lu2017, Zhang2019, Vishina2020, Huber2020}. For example, correlations established with machine learning and data mining methods helped identify new oxides~\cite{Chen2019}, perovskites~\cite{Guistino2016}, and intermetallics~\cite{Miguel-Na5Sn2}. Global structure optimization methods have expanded the exploration beyond known prototypes and resulted in prediction and confirmation of unfamiliar motifs in various materials classes~\cite{ak41, Pickard2011, Oganov2011, Oganov2019, Oganov2019-2}. Unfortunately, the high cost of \textit{ab initio} calculations limits the scope of unconstrained searches. Evaluation of structure stability with less expensive and fairly accurate machine learning potentials (MLPs) has shown great promise for accelerating \textit{ab initio} searches~\cite{ak45} but successful predictions of stable compounds remain scarce~\cite{ak37, Gubaev2019, ak47, Kang2022, PickardJul2022, PickardApr2023, PickardJun2023}. In particular, our recent re-examination of the Li-Sn binary with a MLP has uncovered several stable alloys with large unit cells not detected in \textit{ab initio} searches~\cite{ak47}. In the present study extended to five metal-tin binaries, we aim to demonstrate the applicability and benefit of the developed predictive strategy on a larger scale in a materials class abundant with potential applications.

Tin is a post-transition metal observed in various elemental and multicomponent crystal structure phases~\cite{mehl2021,Felix2011,Jing2016,Wang2011,Cheng2017,Cornelius2017}. At ambient temperature and pressure, 'white tin' crystallizes in the $\beta$-Sn structure and is known as a soft, malleable, and ductile metal. Below 13°C, 'grey tin' adopts the $\alpha$-Sn diamond structure and exhibits a semimetallic behavior. The allotropic $\beta\rightarrow\alpha$ transformation, a so-called 'tin pest' process turning silvery tin objects into grey powder, has slow kinetics due to the high activation energy associated with the change in the atomic coordination from 6 to 4 and the volume expansion by ~27\% ~\cite{Cornelius2017}. Under high pressures and room temperature, tin undergoes a series of transformations to more close-packed structures: $\beta$-Sn$\rightarrow$bct$\rightarrow$bco$\rightarrow$bco+bcc$\rightarrow$bcc at about 11, 32, 40, and 70 GPa, respectively~\cite{Jing2016, Salamat2013}.

The high sensitivity of tin's ground state to the external temperature and pressure conditions can be traced back to the element's particular placement in the periodic table. Within group XIV, tin's position defines the boundary between the covalent bonding for the lighter elements and the metallic bonding for the heavier lead. The propensity of the elements to form covalent bonds can be quantified with a ratio between the $sp^3$ bond formation energy and the $s{\rightarrow}p$ promotion energy cost. The steady 2.8:1.4:1.15:1.02:0.8 decrease of the ratio in the C:Si:Ge:Sn:Pb set~\cite{Methfessel1993} has been linked to the decreasing bond integral strength between the $s$ and $p$ states~\cite{Methfessel1993, Pawlowska1987, Legrain2016}.

The competitiveness of different bonding mechanisms in pure tin contributes to the element's readiness to form alloys. In this respect, tin shares a lot of traits with boron which has been the subject of our past work~\cite{ak28}. This metalloid with three valence electrons also occupies a borderline spot between an insulator (carbon) and a metal (beryllium), assumes several elemental configurations ($\alpha$-B, $\beta$-B, and $\gamma$-B~\cite{Zarechnaya2008, Oganov2009, Barbara2009, Parakhonskiy2011, Kurakevych2012}) due to the frustrated electronic structure, mixes with the majority of metals, and forms extended 2D or 3D covalent networks. The distinction between the two classes is tin's versatility to be either a hosting or an alloying element in compounds which defines the materials' remarkable suite of demonstrated and possible functions. 

Tin's large size and tendency to form extended frameworks give rise to applications as a battery anode material~\cite{Chevrier2011, Wang2018, Wu2022}. Tin alloy anode materials in general are more conductive and safer than graphite-based anodes, and certain tin binary phases have been found to have larger theoretical specific capacities than their commercially-available counterparts (Li$_{22}$Sn$_5$~\cite{Todd2010} and Na$_{15}$Sn$_4$~\cite{Lao2017} have 992 mA h g$^{-1}$ and 847 mA h g$^{-1}$ theoretical specific capacities, respectively, while LiC$_6$ has 372 mA h g$^{-1}$). However, the large volume change of over 250\% (420\%) upon Li (Na) insertion/extraction leads to anode pulverization over just a few cycles~\cite{ZHAO2015, Liu2013, TIAN2015}.

Tin alloys have been extensively investigated as non-toxic alternatives to Pb-free solders and durable joint materials in electronics interconnects~\cite{Wei2008, Ma2009, Ball2012, Cornelius2017, Cheng2017, novakovic2018, BharathKrupaTeja2022, cui2023}. The efforts have focused on finding tin intermetallics that can balance high mechanical stability, high thermal conductivity, resistance to Sn-whisker formation, cost effectiveness, and other factors important for next-generation integrated circuits~\cite{Wei2008,Ball2012}. 

A number of tin-based materials have been studied for their non-trivial topological behavior. They range from pure $\alpha$-Sn~\cite{Fu2007} and Sn-Te/Pb~\cite{Hsieh2012, Dziawa2012, Tanaka2012, Xu2012} 3D materials to atom-thick stanene with a buckled honeycomb morphology~\cite{Felix2011, Liu2011, Zhu2015, Negreira2015}. Nevertheless, a known BaSn$_2$ compound synthesized first almost a decade ago~\cite{Kim2008} received little attention until its potential as a strong TI with a wide 200-meV band gap has been demonstrated in our previous studies~\cite{ak31, ak35}. A recent experimental study provided insights into synthesis and stability of the BaSn$_2$ compound~\cite{Trout2021}.


Given a large body of research dedicated to tin alloys over the past few decades~\cite{Yang2003, Todd2010, Wang2010, Felix2011, LIU2016, Cheng2017, Lao2017}, it is surprising that new ambient-pressure tin-based materials are still being synthesized in common metal-tin binary systems. Recent discoveries include BaSn$_2$ in 2008~\cite{Kim2008}, FeSn$_5$ in 2011~\cite{Wang2011}, and NaSn$_2$ in 2017~\cite{Stratford2017} predicted earlier by our group~\cite{ak31}. It is evident that experimental observation of new phases in this materials class is possible but requires a considerable effort to narrow down the starting compositions and fine-tune the synthesis conditions. 

Our present work involves a systematic exploration of M-Sn (M = Na, Ca, Cu, Pd, and Ag) binaries at different pressures and temperatures. The evolutionary ground state searches based on our neural network (NN) potentials were performed at 0 and 20 GPa. All promising candidates uncovered in the evolutionary searches at $T=0$ K were examined at the DFT level. The thermodynamic stability of select candidates was further investigated in a wide range of temperatures at both NN and DFT levels. As a result, we have identified 11 (18) new intermetallics expected to be synthesizable at ambient (high) pressure. The rest of the manuscript is organized as follows. Section~\ref{methods} provides details about the DFT calculations, the construction of the NN potentials, and our compound prediction strategy. Section~\ref{results} overviews prior work on the five M-Sn binaries, highlights their key applications, and presents our findings on new stable alloys. Section~\ref{summary} summarizes our results and discusses remaining challenges of materials prediction.

\section{Methods}
\label{methods}

\subsection{Density functional theory calculations} 
All DFT calculations were
performed with the Vienna \textit{ab-initio} simulation package ({\small
VASP})~\cite{VASP,VASP2,VASP3,VASP4}. We chose projector augmented wave potentials~\cite{PAW}
with semi-core electrons: Na\_{\it pv}, Ca\_{\it sv}, Cu\_{\it pv}, Pd\_{\it pv}, Ag, and Sn\_{\it d}. Unless specified otherwise, we
used the Perdue-Burke-Ernzerhof (PBE) exchange-correlation functional~\cite{PBE}
within the generalized gradient approximation (GGA)~\cite{Langreth1983} and the
energy cutoff of 500 eV. Select phases were examined within the local density approximation (LDA)~\cite{LDA1,LDA2} or with the strongly-constrained and appropriately-normed (SCAN) functional in the meta-GGA~\cite{SCAN}. All crystalline structures were evaluated with dense ($\Delta k
\sim 2\pi \times 0.02$ $\si{\angstrom}^{-1}$) Monkhorst-Pack $k$-point meshes~\cite{Monkhorst1976}.

\subsection{Neural network interatomic potentials} 
The MLPs employed in this study rely on the Behler-Parrinello (BP) descriptor for converting local atomic environments into NN input signals~\cite{behler2007}. The parameter sets defining the BP pair and triplet symmetry functions within an extended 7.5 \si{\angstrom} cutoff radius can be found in our previous studies~\cite{ak40, ak41}. The NN models were constructed using an automated iterative scheme implemented in our {\small MAISE-NET} framework~\cite{ak41} that features an evolutionary sampling algorithm to generate representative reference structures and a stratified training protocol to build multicomponent models on top of elemental ones~\cite{ak34}. The resulting datasets generated in four {\small MAISE-NET} cycles consist of (i) partially optimized crystal structures with up to 10 atoms for elements and 32 atoms for binaries in the $0.125 \leq x \leq 0.875$ composition range obtained in short evolutionary runs at 0, 10, 30, and 50 GPa and (ii) a small fraction (10-20\%) of compressed and expanded configurations of basic prototypes and found minima. About 10\% of structures randomly selected from the former group were set aside for testing.

In the stratified parameterization of models for the Sn binary systems, we relied on our previously developed Na, Ca, Cu, Pd, Ag, and Sn NNs~\cite{ak41} with the 51-10-10-1 architecture and 641 adjustable parameters (the total of $(8+43+1)\times 10 + (10+1)\times (10) + (10+1)\times 1$ weights corresponding to 8 BP pair functions, 43 BP triplet functions, and 1 bias in the input layer, 10 neurons and 1 bias in each of the two hidden layers, and 1 output). The binary M-Sn NNs with a 145-10-10-1 architecture and 1,880 adjustable parameters (the total of $(8_{\text{MSn}}+43_{\text{MMSn}}+43_{\text{MSnSn}}) \times 10_{\text{M}} + (8_{\text{SnM}}+43_{\text{SnSnM}}+43_{\text{SnMM}}) \times 10_{\text{Sn}}$ interspecies weights connecting the input symmetry functions with 10 neurons in the first hidden layer of each elemental NN) were fitted to datasets of only binary structures.

Information about the dataset sizes, NN root-mean-square errors, and error distributions is given in Table~\ref{tab:models} and Figs. S1-S5. As discussed in our previous studies~\cite{ak41, ak47}, the relatively modest accuracy of about 10 meV/atom results from the inclusion of unfamiliar configurations identified during NN-based structure search test runs at the end of each {\small MAISE-NET} cycle. The addition of diverse high-energy configurations generally increases the total error by about a factor of two but greatly reduces the number of artificial minima. This strategy has been found to ensure a desired balance between accuracy and robustness for the intended NN application in global structure searches. The Na-Sn system proved to be the most challenging binary to model because the experimentally observed ground states have large unit cell sizes outside our standard sampled range and the NN could not learn how to accurately describe the exotic Sn frameworks~\cite{ak31}. Incorporation of the equation-of-state data for the nine structures into the training set resulted in only a marginal NN improvement, as the original 19-96 meV/atom errors decreased to 4-68 meV/atom for this subset.

Considering that this study focuses on Sn alloys, we examined the description of pure Sn in more detail. Fig. S6 displays the relative stability of relevant Sn phases in the 0-20 GPa range evaluated with the reference PBE flavor, our NN potential, and the most accurate modified embedded atom model (MEAM) developed for this element~\cite{ko2018}. First, it is worth reviewing how the DFT predictions agree with experimental observations. The likely overestimation of the $\Delta E_{\beta-\alpha}$ energy difference in common DFT approximations along with the limitations of the standard harmonic approximation for evaluating free energies leads to notably higher estimates of the $T_{\alpha \rightarrow \beta}$ transition temperature~\cite{Legrain2016,ko2018,mehl2021,nitol2023}, 540 K in our case versus measured 286 K~\cite{Pavone1998}. This discrepancy makes it difficult to make definitive predictions about possible high-$T$ ground states at the Sn-rich end of the phase diagram. The PBE results at $T=0$ K also indicate stability of hcp-Sn in the 5.5-15 GPa window not detected experimentally and favorability of bco over bct up to about 20 GPa in disagreement with reported data~\cite{Salamat2013}. Our NN model reproduces $\Delta E_{\beta-\alpha}$ within 3.5 meV/atom and predicts $T_{\alpha\rightarrow\beta}=530$ K but does not differentiate bco, bct, and bcc above 14 GPa. Overall, the NN offers a satisfactory description of relative stabilities in the considered pressure range given that all structures were fully relaxed with the corresponding methods and that the absolute error doubles in the calculation of enthalpy differences between two phases. The MEAM agrees well with DFT describing phases at ambient pressure. Since it was not parameterized to model compressed Sn configurations, it is not unexpected to see its less accurate performance resolving competing phases under high pressures. 
At the same time, the MEAM significantly disfavors the viable but elusive $\gamma$-Sn with the simple hexagonal structure~\cite{mehl2021} and our MEAM-based evolutionary searches uncovered an artificial ground state (oI4-$Immm$, $a=3.2485$ \AA, $b=4.5366$ \AA, $c=8.3219$ \AA, and a single $4j$ (1/2,0, 0.3276) Wyckoff position) 11 meV/atom below $\alpha$-Sn at 0 K and 0 GPa. Finally, an interesting hybrid model, a combination of EAM and rapid artificial NN potential, has been recently developed and demonstrated to have a much-improved description of Sn phases in a wide range of temperatures and pressures~\cite{nitol2023}.

\newcolumntype{C}{>{\centering\arraybackslash}X}
\newcolumntype{R}{>{\raggedleft\arraybackslash}X}

\begin{table}[!b] 
\begin{tabularx}{0.45\textwidth} {r r r | c c c c c c } 
  \hline \hline 
  & & & \multicolumn{2}{c} {Training set} & \multicolumn{2}{c} {Testing set} & \multicolumn{2}{c} {Testing errors} \\ 
  & & & $E$ & $F$ & $E$ & $F$ & $E$ & $F$ \\ 
  & & & size & size & size & size &  meV/at. & meV/\si{\angstrom} \\  
  \hline 
  & Na & & 4584 & 34059 & 509 & 3882 & 0.7 & 3 \\
  & Ca & & 4483 & 33327 & 498 & 4011 & 6.2 & 30 \\
  & Cu & & 4316 & 32199 & 479 & 4053 & 1.8 & 10 \\
  & Pd & & 4398 & 33225 & 488 & 4101 & 4.6 & 40 \\
  & Ag & & 4421 & 35889 & 491 & 4416 & 1.9 & 11 \\
  & Sn & & 4346 & 29694 & 482 & 3687 & 8.3 & 35 \\
  & Na-Sn & & 6953 & 33501 & 772 & 2610 & 12.5 & 43 \\ 
  & Ca-Sn & & 7211 & 29376 & 801 & 3327 & 14.5 & 65 \\ 
  & Cu-Sn & & 7685 & 69579 & 853 & 7578 & 9.5 & 45 \\ 
  & Pd-Sn & & 5507 & 46272 & 611 & 4812 & 9.6 & 51 \\ 
  & Ag-Sn & & 5410 & 37548 & 601 & 4236 & 8.4 & 33 \\ 
  \hline \hline  
  \end{tabularx} 
  \caption{Dataset sizes and root-mean-square errors of constituent elemental and combined binary NN models.} 
\label{tab:models} 
\end{table}

\subsection{Evolutionary structure searches} 
Global structure optimizations were performed for each binary system with our {\small MAISE} package~\cite{ak41}. Evolutionary searches were carried out for selected fixed M$_{1-x}$Sn$_x$ compositions with $0.125 \leq x \leq 0.875$. We considered up to 8 formula units and limited the structure sizes to 24 atoms per unit cell. Randomly generated populations of 32 members were evolved for up to 168 generations with standard evolutionary operations. The number of generations was set to $g=N + N^2/4$ as a function of the unit cell size that has been found to be usually sufficient to reach search convergence in our previous study on Li-Sn~\cite{ak47}. Evolutionary operations consisted of mutations of 4 single parents (random atom displacements, atom swaps, and unit cell distortions), injections of 4 new random structures, and crossovers of 24 pairs of parents (combination of two roughly equal parts obtained with planar cuts)~\cite{ak16,ak41}. Child structures were locally relaxed with the NN potentials for up to 300 Broyden–Fletcher–Goldfarb–Shanno minimization steps and assigned a fitness based on the final enthalpy. Our fingerprint method based on the radial distribution function (RDF)~\cite{ak16,ak23,ak41} was used to identify and eliminate similar structures.

For select compositions, we carried out additional searches with our hybrid NN+DFT approach that ensured more reliable convergence to putative ground states in the study of Au nanoparticles~\cite{ak40}. The method involves structural relaxations with a NN model followed by static DFT calculations to better resolve candidate's favorability during evolutionary runs. None of the 21 hybrid searches completed for the Na-Sn and Pd-Sn binaries improved on the best candidates obtained with the standard NN-based runs.

\subsection{Vibrational property analysis} 
Most of the phonon calculations were performed in the harmonic approximation with the finite displacement method as implemented in Phonopy~\cite{phonopy}. We used symmetry-preserving expansions of primitive or conventional unit cells to generate supercells with 72-256 atoms and applied 0.1 \si{\angstrom} displacements. The Gibbs free energy corrections due to the vibrational entropy were included via summation over $20\times20\times20$ grids in the Brillouin zone approximating the integral $\Delta F_{\text{vib.}} = k_b T \int_{0}^{\infty} d\omega g(\omega) ln[2sinh(\hbar\omega/2k_b T)]$. 

In some cases, we performed quasi-harmonic approximation (QHA) calculations to determine the significance of anharmonic effects at elevated temperatures. For a given structure, the procedure involves (i) creation of a uniform grid of expanded and compressed volumes about the equilibrium; (ii) relaxation of each structure with VASP under the constant-volume constraint; (iii) phonon calculation in the harmonic approximation at each volume using Phonopy and VASP; (iv) fitting the free energy points at each temperature with a third-order polynomial; and (v) finding the minimum of the free energy as a function of volume at each temperature.

\subsection{Post-search stability analysis} 

The protocol of selecting viable candidate phases found in NN-based evolutionary searches for further examination with DFT approximations was detailed in our previous studies~\cite{ak40,ak47}. For creating pools of possible $T=0$ K ground states, we adopted similar enthalpy windows of up to 20 meV/atom, approximately double of the NN model errors, to include distinct minima above the lowest-enthalpy configuration obtained in each evolutionary run. We used the same 0.95 RDF dot product cutoff to exclude similar structures and a 0.1 \AA tolerance to symmetrize each unit cell before their re-optimization at the DFT-PBE level~\cite{ak47}. For making and evaluating pools of possible high-$T$ ground states, we adhered to the following steps designed to reduce the error and computational time~\cite{ak47}: (i) calculate $\Delta F^{\textrm{NN}}_{\textrm{vib.}}$ for all candidate structures in the $T=0$ K pools within the 20 meV/atom DFT enthalpy window; (ii) construct a convex hull at 600 K using $\Delta G^{\textrm{DFT+NN}} = \Delta H^{\textrm{DFT}} + \Delta F^{\textrm{NN}}_{\textrm{vib.}}$ and select all structures with Gibbs free formation energies no further than 10 meV/atom from the tie-line; and (iii) evaluate thermal corrections for these structures with DFT and determine high-$T$ ground states via convex hull construction at every 10 K up to 2000 K.

\begin{table}[!t] 
\begin{tabularx}{0.45\textwidth} {r r r | C C C C } 
  \hline \hline 
  & & & Pool size & $\Delta H^{\textrm{DFT}}_{\textrm{relax}}$ & Pool size & $\Delta F^{\textrm{NN-DFT}}_{\textrm{vib.}}$ \\
  & & & at 0 K & meV/at. & at high $T$ & meV/at. \\
  \hline 
  & Na-Sn & & 3370 & 6.9 & 43 & 1.4 \\ 
  & Ca-Sn & & 1780 & 4.2 & 25 & 1.4 \\ 
  & Cu-Sn & & 2590 & 1.7 & 27 & 1.3 \\ 
  & Pd-Sn & & 1539 & 3.4 & 39 & 0.8 \\ 
  & Ag-Sn & & 4602 & 1.5 & 39 & 0.5 \\ 
  \hline \hline  
  \end{tabularx} 
  \caption{Correspondence between NN and DFT results evaluated for pools of NN minima selected at zero and high temperatures. The $\Delta H^{\textrm{DFT}}_{\textrm{relax}}$ values represent average drops in enthalpy upon re-optimization of NN minima with DFT at $T=0$ K. The $\Delta F^{\textrm{NN-DFT}}_{\textrm{vib.}}$ values are average deviations between NN and DFT relative free energy corrections from vibrational entropy at $T=600$ K (see text for more detail).}
\label{tab:pool} 
\end{table}

Table~\ref{tab:pool} presents information on the pool sizes and the correspondence between NN and DFT results for each M-Sn binary. The relatively low $\Delta H^{\textrm{DFT}}_{\textrm{relax}}$  values of the average enthalpy drops during DFT re-optimizations, ranging from 1.5 meV/atom in Ag-Sn to 6.9 meV/atom in Na-Sn, demonstrate the proximity of the starting configurations in the NN pools to the corresponding DFT minima. This observation lends further support to the aforementioned NN+DFT hybrid method as an effective approach of predicting the DFT ranking for NN candidate phases without DFT relaxations at $T=0$ K. To quantify the agreement between the NN and DFT methods for assessing vibrational entropy corrections at elevated temperatures, we determined average deviations $\Delta F^{\textrm{NN-DFT}}_{\textrm{vib.}}=\Delta F^{\textrm{NN}}_{\textrm{vib.}}-\Delta F^{\textrm{DFT}}_{\textrm{vib.}}$. The free energy differences at the two levels of theory were calculated separately for each pool with respect to the pool’s mean free energy and then averaged, as root-mean-square, over all pools for each binary. The low deviation values of less than 1.4 meV/atom justify the use of the NN-based prescreening as a means of finding viable high-$T$ ground states.


\section{Results}
\label{results}

%
%

\subsection{The Na-Sn binary}

Investigation of the compound-rich Na-Sn binary system extends over a century. In 1920, the existence of four alloys at the 4:1, 2:1, 1:1, and 1:2 compositions was demonstrated with electrochemical experiments~\cite{Kremann1920}. Further synthesis and characterization work uncovered complex phases at or near the 15:4, 5:2, 7:3, 9:4, 7:12, 5:13, 1:3, 1:4, and 1:5 stoichiometries~\cite{Sangster1998}. The recent research on these alloys has been motivated primarily by their potential use in energy storage applications based on abundant and inexpensive sodium~\cite{Chevrier2011, Wu2022}. As the graphite anode materials utilized in commercial lithium-ion batteries are not suitable for storing larger sodium ions, Na-Sn alloys have been considered as promising anode alternatives with a high theoretical capacity of about 847 mA h g$^{-1}$ when converting Sn into Na$_{15}$Sn$_4$~\cite{Stratford2017}. A new layered NaSn$_2$ material was also predicted to be a topologically nontrivial metal~\cite{ak31}. Stratford {\it et al.}~\cite{Stratford2017} detected this phase and proposed solutions to previously unidentified intermediates in their comprehensive experimental and DFT study of the full Na-Sn composition range.

\begin{figure}[t] \centering
  \includegraphics[width=0.476\textwidth]{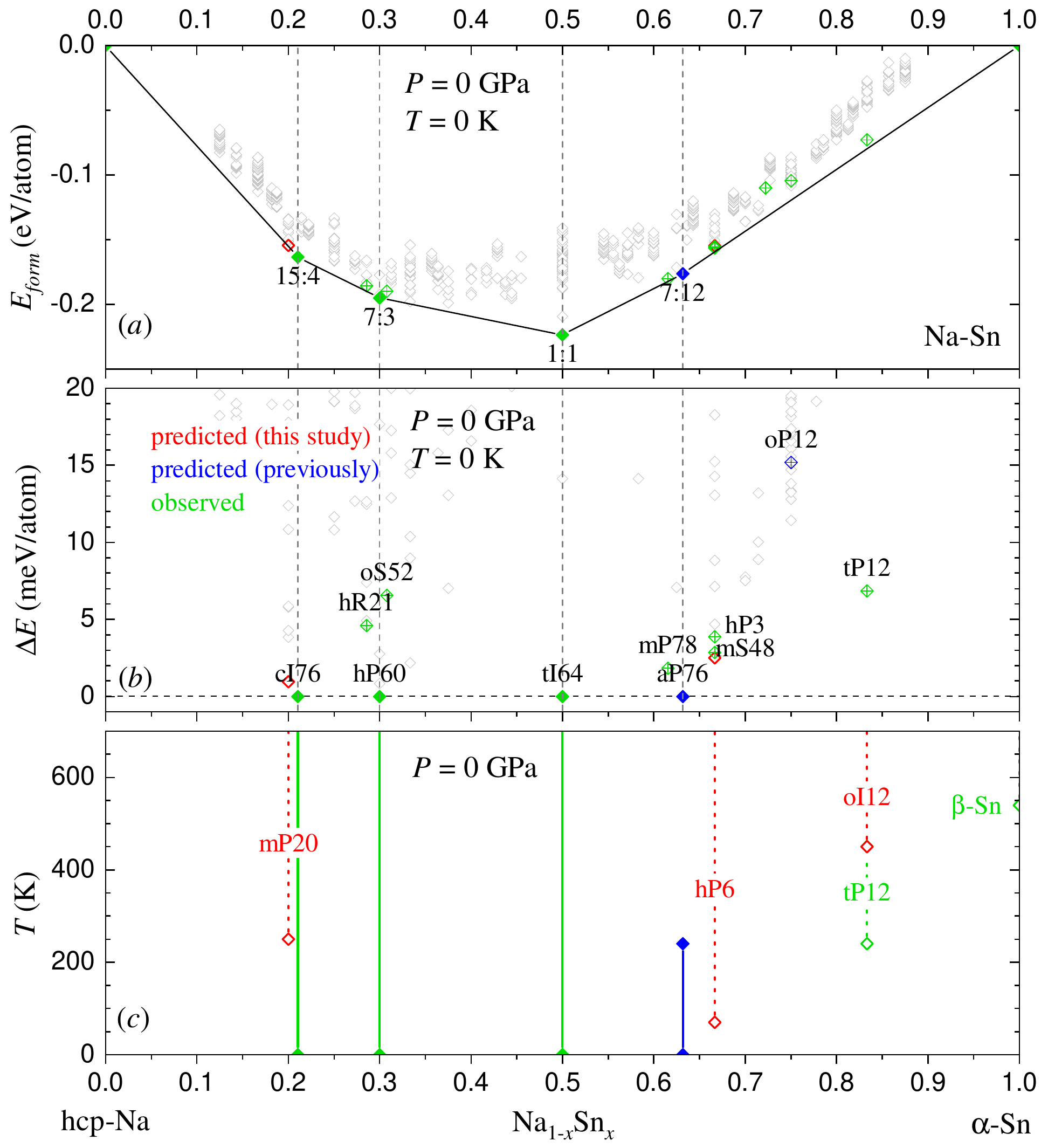} 
  \caption{Calculated stability of Na-Sn intermetallics at ambient pressure. The three panels show (a) the convex hull constructed at $T=0$ K, (b) energy distances to the tie-line at $T=0$ K, and (c) temperature ranges of thermodynamic stability. The solid, hollow, and crossed diamonds correspond to phases found in our calculations to be stable at $0$ K, stabilizing at elevated temperatures, and metastable at all considered temperatures, respectively. The green, blue, and red colors denote experimentally observed, previously considered, and our predicted phases.}
\label{fig:NaSnhull00} 
\end{figure}

\begin{figure}[t] \centering
  \includegraphics[width=0.48\textwidth]{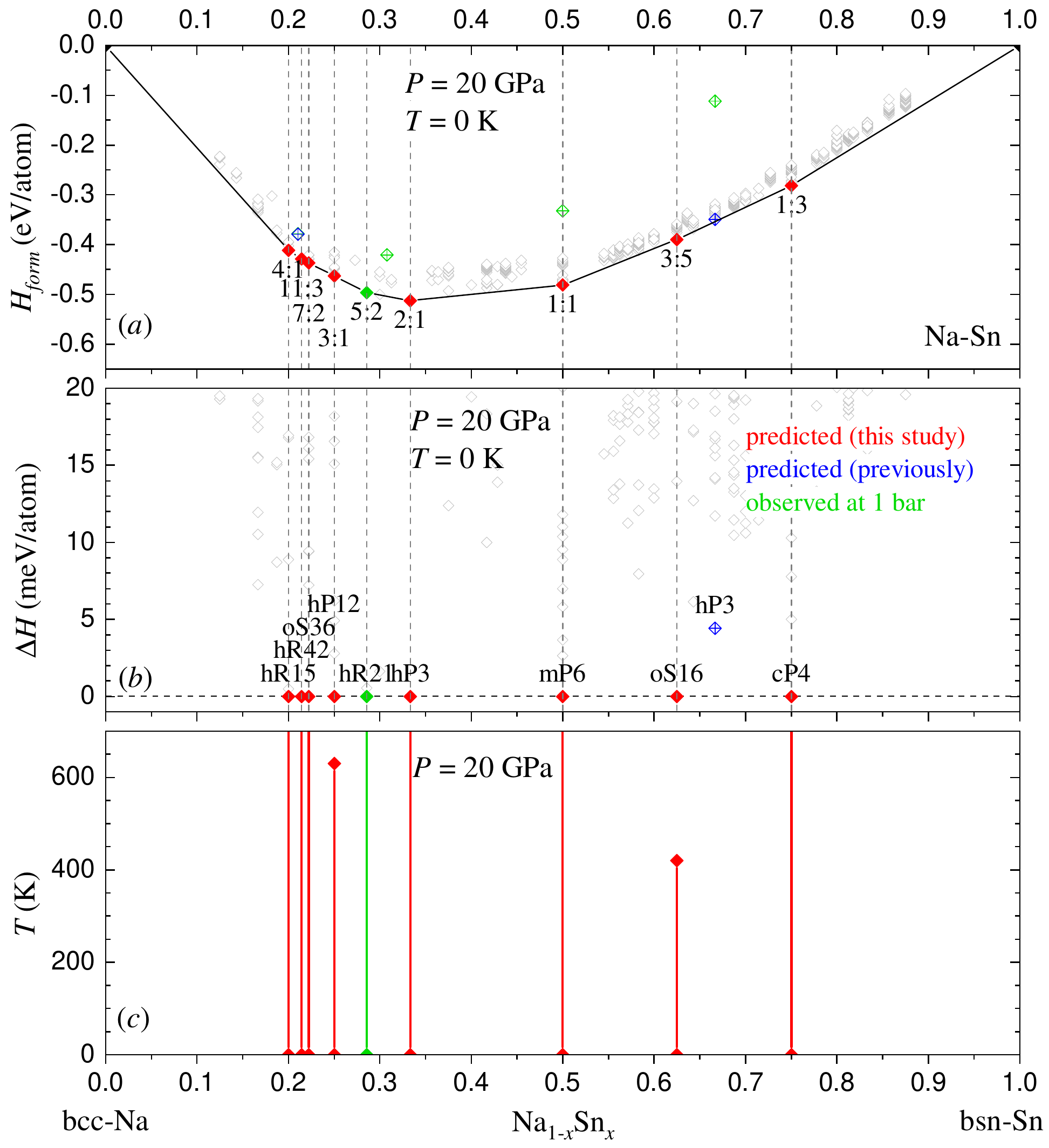} 
  \caption{Calculated stability of Na-Sn intermetallics at 20 GPa. The three panels show (a) the convex hull constructed at $T=0$ K, (b) energy distances to the tie-line at $T=0$ K, and (c) temperature ranges of thermodynamic stability. The symbol and line styles are the same as in Fig.~\ref{fig:NaSnhull00}.} 
\label{fig:NaSnhull20} 
\end{figure}

The presence of phases with large unit cells, partial occupancies, and unique morphologies made the identification of the full set with unconstrained structure searches impractical. Following the strategy used in our Li-Sn study, we performed the thermodynamic stability analysis of the Na-Sn binary by combining the previously observed phases and low-enthalpy ones found in our NN-based evolutionary searches (see Figs.~\ref{fig:NaSnhull00} and~\ref{fig:NaSnhull20}). At ambient pressure and zero temperature, the resulting convex hull is comprised of the cI76-Na$_{15}$Sn$_4$~\cite{Volk-Na9Sn4-Na15Sn4}, hP60-Na$_7$Sn$_3$~\cite{Baggeto-Na7Sn3}, and tI64-NaSn~\cite{Volk-NaSn} ordered ground states and an aP76-Na$_7$Sn$_{12}$ representation~\cite{ak31} of the disordered monoclinic phase~\cite{Fassler-Na7Sn12}. As discussed in our previous study~\cite{ak31} and below, the related ordered mS48-NaSn$_2$ and mP78-Na$_5$Sn$_8$ variants~\cite{Stratford2017} are actually metastable by 2.6 and 1.8 meV/atom, respectively. The known hR21-Na$_5$Sn$_2$~\cite{Obrovac-Na5Sn2, Baggeto-Na5Sn2}, oS52-Na$_9$Sn$_4$~\cite{Volk-Na9Sn4-Na15Sn4}, and oS288-Na$_5$Sn$_{13}$~\cite{Vaughey1997} phases hover about 5, 7, and 23 meV/atom above the corresponding tie-lines. 

\begin{figure}[t] \centering
  \includegraphics[width=0.48\textwidth]{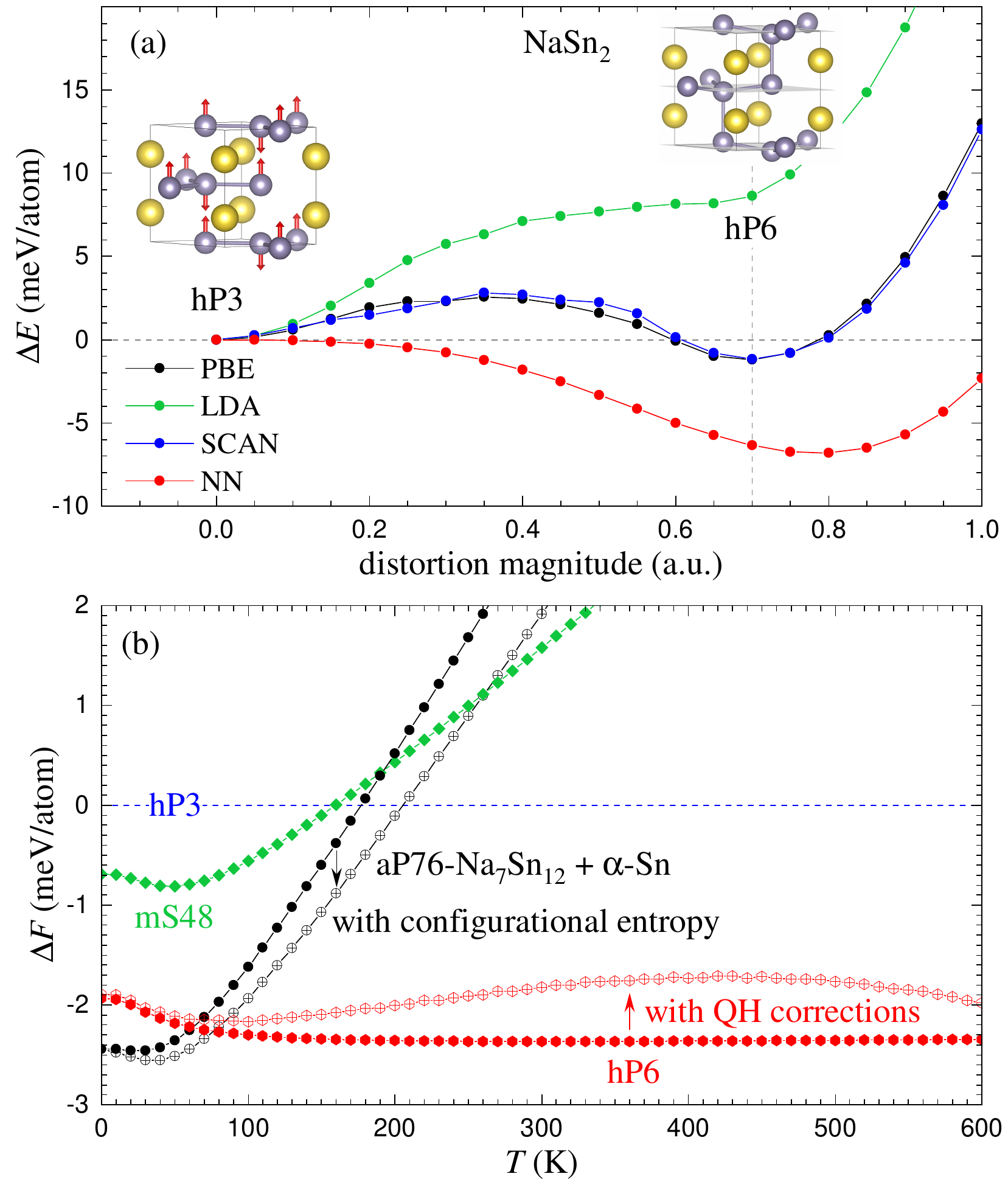} 
  \caption{Stability analysis of NaSn$_2$ phases. (a) Relative energy of the hexagonal NaSn$_2$ structures along the hP3 to hP6 transformation path following an out-of-plane Sn phonon mode. At each point, the magnitude of the Sn layer distortion was kept fixed while the unit cell shape was optimized. (b) Free energies of NaSn$_2$ polymorphs relative to hP3 as a function of temperature. The solid points correspond to free energies with vibrational entropy contributions calculated in the harmonic approximation. The crossed red hexagons show the significance of the quasi-harmonic (QH) corrections on the relative stability between hP6 and hP3. The crossed black circles illustrate the effect of the configurational entropy in the disordered aP76-Na$_7$Sn$_{12}$ on the distance to the aP76-Na$_7$Sn$_{12}\leftrightarrow\alpha$-Sn tie-line.}
\label{fig:NaSnDISP} 
\end{figure}

Our screening for high-$T$ ground states identified three new viable candidates, mP20-Na$_4$Sn, oI12-NaSn$_5$, and hP6-NaSn$_2$, that are within 0.4 meV/atom, 32.6 meV/atom, and 2.8 meV/atom of stability at 0 K in our PBE calculations (see Table S1). With vibrational entropy corrections included, mP20-Na$_4$Sn stabilizes at 250 K and lies 1.7 meV/atom below the hcp-Na$\leftrightarrow$cI76-Na$_{15}$Sn$_{4}$ tie-line at 650 K, which is the approximate melting temperature at this Na-Sn composition~\cite{Sangster1998}. The experimentally observed tP12-NaSn$_5$ phase~\cite{Fassler-NaSn5} is stabilized at 240 K (note that the choice of the metastable $\beta$-Sn in the calculation of formation energies makes this alloy stable at $T=0$ K~\cite{Stratford2017}). Our proposed oI12-NaSn$_5$ polymorph becomes the ground state at this stoichiometry above 450 K. We did not attempt to calculate phonons for the metastable oS288-Na$_5$Sn$_{13}$ or stable aP76-Na$_7$Sn$_{12}$ due to their exceptionally large sizes and/or low symmetry. We did approximate the free energy of the latter by adding a linear interpolation of the vibrational entropy terms evaluated for the related mS48-NaSn$_2$ and mP78-Na$_5$Sn$_8$ at the adjacent compositions. This allowed us to estimate that the new hP6-NaSn$_2$ phase should become the ground state at 70 K and, in turn, destabilize aP76-Na$_7$Sn$_{12}$ above 240 K. The configuration entropy contribution in Na$_7$Sn$_{12}$ is expected to be insignificant because only $M=4$ out of 30 Na atom sites in the $N=78$-atom unit cell have fractional occupancies. As shown in Fig.~\ref{fig:NaSnDISP}(b), the $\frac{M/N}{1-x_dM/N}kT[x_d\ln(x_d)+(1-x_d)\ln(1-x_d)]= -kT/19 \ln(2)$ correction evaluated for half-filled $4g$ sites ($x_d=0.5$) in the assumption that all configurations are equiprobable shifts the hP6 stabilization temperature up by just $\sim 15$ K.

\begin{figure}[t] \centering
  \includegraphics[width=0.48\textwidth]{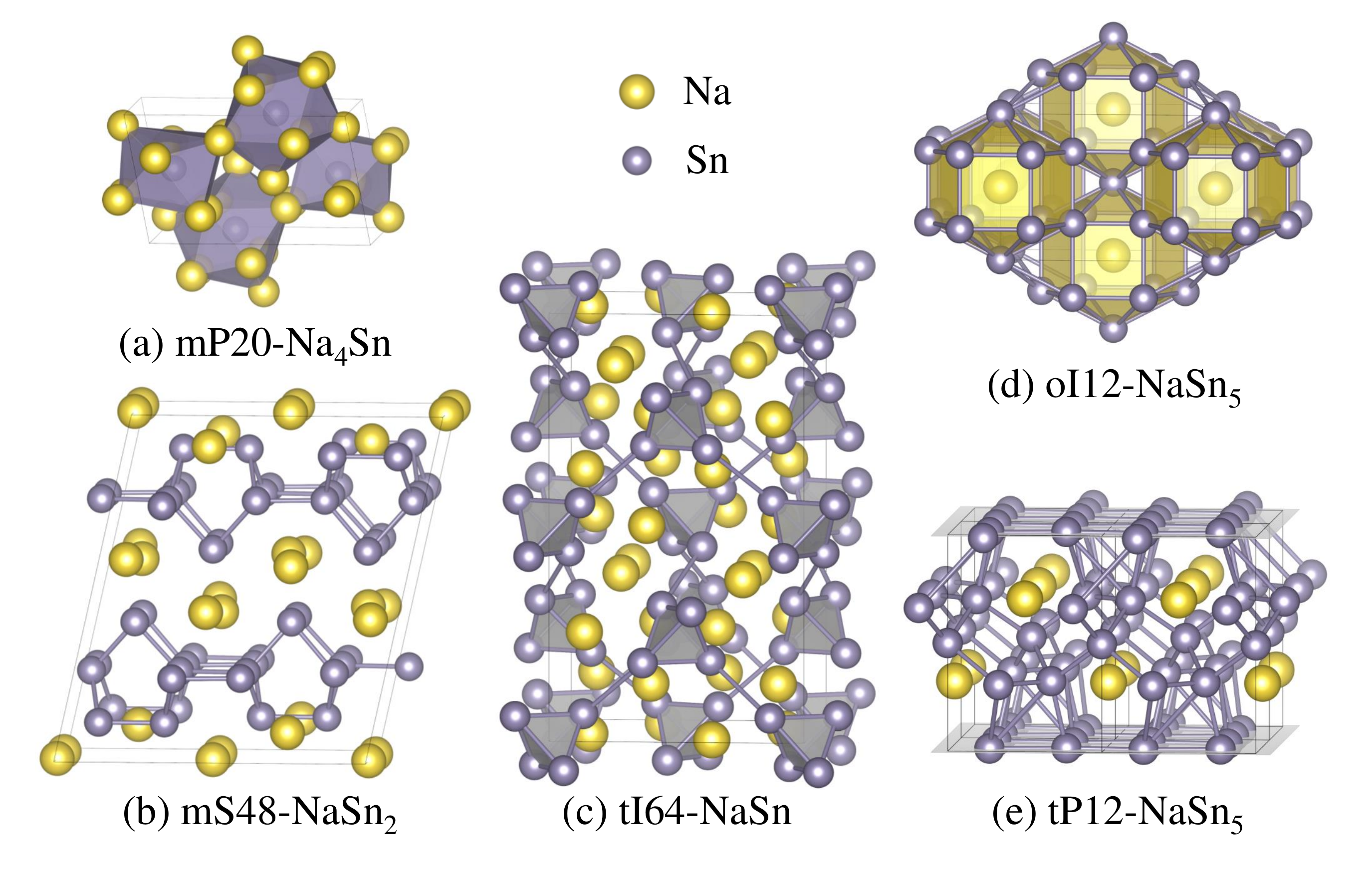} 
  \caption{Structures of select Na-Sn phases. The polyhedra in (a,d) show the local environments of the minority species in the predicted high-temperature oS20-Na$_4$Sn and oI12-NaSn$_5$ ground states, respectively.} 
\label{fig:NaSnPics} 
\end{figure}

It has been discussed that the Na-Sn and related Li-Si/Ge/Sn binaries exhibit similar morphological trends, as different intercalated covalent frameworks appearing in Sn-rich alloys give way to Sn-Sn dimers and eventually to isolated Sn atoms with the increase of the alkali-metal concentration~\cite{Morris2014,Jung2015,Stratford2017,ak31}. The tP12-NaSn$_5$ phase features a complex 3D framework with square Sn nets crosslinked by a percolating web of Sn bonds, from 4 to 7 per atom, that has well-defined channels conducive for Na migration (see Fig.~\ref{fig:NaSnPics}). In fact, Stratford {\it et al.}~\cite{Stratford2017} identified metastable NaSn$_3$ and NaSn$_4$ phases with similar 3D morphologies, which could explain the reported but not identified alloy at the latter composition~\cite{Sangster1998}. Our predicted oI12-NaSn$_5$ is comprised of fused elongated hexagonal bipyramids that effectively trap Na ions. In tI64-NaSn, the Sn framework consisting of tetrahedra about 3.70 \AA\ apart was found with a combination of \textit{ab initio} molecular dynamics and a reverse Monte Carlo refinement to be described better with an amorphous structure~\cite{Stratford2017}. The predicted Na-rich mP20-Na$_4$Sn phase has a fairly uniform distribution of Sn atoms with Na$_{9}$ and Na$_{11}$ local coordinations.

The phases at and around the 1:2 composition deserve a closer look. It has been discussed previously that the mS48-NaSn$_2$, aP76-Na$_7$Sn$_{12}$, and mP78-Na$_5$Sn$_8$ phases are closely related and can be obtained by different population of the available Na sites between disjointed polyanion Sn layers stacked along the $c$-axis~\cite{ak31}. The most stable aP76 structure was obtained from mP78 by removing two $4g$ Na atoms that results in the largest separation between the vacancies. Our original motivation for studying NaSn$_2$ was to examine whether the compound could crystalize in the AlB$_2$ structure with weakly interacting honeycomb layers and be subsequently exfoliated into stanene. We found that the phase does stabilize over the known phases at elevated temperatures and pressures but owes its stability to strong interlayer covalent bonds that prevent the material from being a good precursor for stanene synthesis. The following {\it ab initio} analysis by Stratford {\it et al.}~\cite{Stratford2017} supported our conclusions regarding the hP3-NaSn$_2$ stability and showed that a derived metastable NaSn$_3$ phase with partial substitution of Na for Sn is consistent with experimental observations.

The new hP6-NaSn$_2$ ground state identified in our unconstrained searches has an unexpectedly simple CaIn$_2$ prototype. It can be constructed from hP3 by doubling the unit cell and distorting the Sn honeycomb framework in and out of the basal plane. Our examination of the transformation path along the corresponding phonon mode in Fig.~\ref{fig:NaSnDISP} shows that both hexagonal structures represent local minima at the PBE and SCAN levels, which is further supported with a phonon dispersion analysis that indicates their dynamical stability (see Fig. S7). The presence and depth of the hP6 minimum do depend on the level of theory. Namely, previous PBE calculations revealed a slow convergence of the out-of-plane phonon mode in hP3~\cite{Stratford2017}, our NN model predicts a 7 meV/atom drop along the barrier-free transition path, while the LDA produces a plateau 7 meV/atom above the starting configuration. Our further PBE calculations demonstrated that thermodynamic corrections evaluated in the QHA have little effect on the relative free energy (see Fig.~\ref{fig:NaSnDISP}(b) and Fig. S8). Hydrostatic compression, on the other hand, disfavors the hP6 minimum at 0.4 GPa and the distorted structure relaxes back to hP3 above 2 GPa (see Fig. S9). If the proposed phase does form it should be easily distinguishable from hP3 because the distortion induces noticeable changes in structural and vibrational properties. The contraction of the $a$ lattice constant by 2.4\% and the expansion of the $c$-axis by 9.1\% results in considerable shifts of powder XRD peak positions (see Fig. S10). The change in local Sn atom environments from three 3.102-\AA\ in-plane and two 3.255-\AA\ out-of-plane bonds in hP3 to one out-of-plane 3.056-\AA\ and three 3.071-\AA\ bonds in hP6 softens the highest optical modes by 17\%.

At 20 GPa pressure, we find that the convex hull is defined by nine phases (Fig.~\ref{fig:NaSnhull20}). Some of the ambient-pressure ground states, {\it e.g.}, tI64-NaSn and mS48-NaSn$_2$, destabilize by more than 0.1 eV/atom and only one, hR21-Na$_5$Sn$_2$, remains stable under this pressure. The eight new ground states found in our searches exhibit compact morphologies more favorable under compression. On the Na-rich side, the hR15-Na$_4$Sn, hR42-Na$_{11}$Sn$_3$, hP12-Na$_3$Sn, and hR21-Na$_5$Sn$_2$ phases with large $c/a$ ratios resemble the Li-rich bcc alloys with different stacking sequences along the [111] direction investigated in our previous study~\cite{ak47}. According to our notation that specifies the separation between Sn layers along the $c$ axis~\cite{ak47}, these structures can be described simply as $|5|$, $|455|$, $|345|$, and $|34|$, respectively. In the middle of the composition range, the mP6-NaSn phase isostructural to mP6-LiSn~\cite{Muller1973} is a different bcc-based alloy that exhibits zig-zag Sn nets. At the Sn-rich end, oS16-Na$_3$Sn$_5$ displays square Sn nets bridged by rows of Sn atoms and a cP4-NaSn$_3$ phase has the common L$_{12}$ prototype. The observed conformity between several compressed Na-Sn and ambient-conditions Li-Sn ground state structures indicates the similarity of effective size ratios between the alkali metal and tin ions under the corresponding 20 GPa and 1 bar pressures.

%
%

\subsection{The Ca-Sn binary}

Early work on the Ca-Sn binary materials indicated the existence of oP12-Ca$_2$Sn, oS8-CaSn, cP4-CaSn$_3$, and tI204-Ca$_{31}$Sn$_{20}$ phases~\cite{Hansen1958, Fornasini1977, Massalski1990}. In 2000, Palenzona and Fornasini~\cite{PALENZONA2000} systematically re-examined the binary system in the full composition range with a combination of differential thermal analysis, metallographic analysis, and single-crystal and powder X-ray diffraction. The study led to the discovery of new oP52-Ca$_7$Sn$_6$, tP118-Ca$_{36}$Sn$_{23}$, and tI32-Ca$_{5}$Sn$_{3}$ intermetallics and the compilation of the most complete phase diagram to date.

\begin{figure}[b] \centering
  \includegraphics[width=0.48\textwidth]{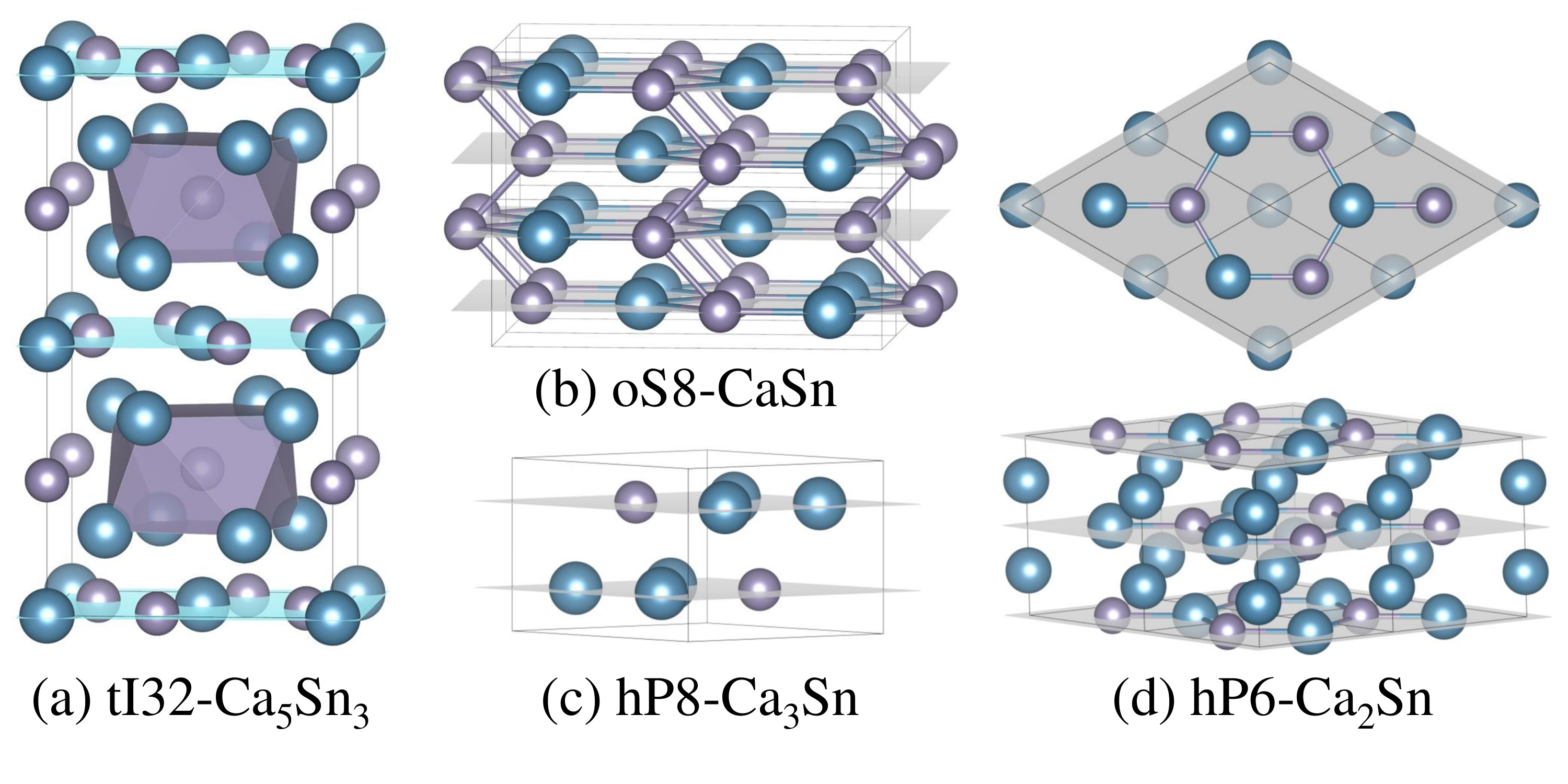} 
  \caption{Select structures of Ca-Sn phases observed under ambient pressure (a,b) and predicted to be stable at 20 GPa (c,d). The Ca and Sn atoms are shown as large and small spheres, respectively. The tI32 structure has pure Sn layers shown with coordination polyhedra and mixed Sn-Ca layers shown with shaded planes. The oS8 structure features stretched honeycomb layers linked via zig-zag Sn-Sn bonds.} 
\label{fig:CaSn-pic} 
\end{figure}

Structural, bonding, electronic, topological, and other properties of Ca-Sn compounds have been analyzed in a number of detailed DFT studies. Ohno \textit{et al.}~\cite{OHNO2006} used a combination of DFT and CALPHAD methods to obtain the thermodynamic model of the binary system. Yang \textit{et al.}~\cite{YANG2010} compared {\it ab initio} and previously measured heats of formation, constructed the convex hull, and analyzed the intermetallics’ electronic and elastic properties. Engelkemier \textit{et al.}~\cite{Engelkemier2013} used their DFT-chemical pressure approach to rationalize how the atomic sizes determine the favorability of Ca-Sn superstructures derived from the W$_5$Cr$_3$ prototype. A recent demonstration of the Ca$_7$Sn$_6$ function as an anode material with a high working voltage of 4.45 V, excellent cyclability with 95\% retention after 350 cycles, and good capacity of 85 mA h g$^{-1}$ for Ca-ion batteries~\cite{Wang2018} spurred computational studies of the binary compounds’ electrochemical, stability, and elastic properties~\cite{Yao2019,Woodcox2021} and further experimental investigations of the Ca-Sn materials’ energy storage potential~\cite{Zhao2022}. CaSn$_3$ with the Cu$_3$Au-type (L1$_2$) structure has been shown to have nontrivial topological and superconducting properties~\cite{Luo2015,Gupta2017,Zhu2019, Siddiquee2021, Siddiquee2022}, while CaSn has been found to be a nodal-line semimetal with potential for topological superconductivity~\cite{Xu2021}.

\begin{figure}[t] \centering
  \includegraphics[width=0.476\textwidth]{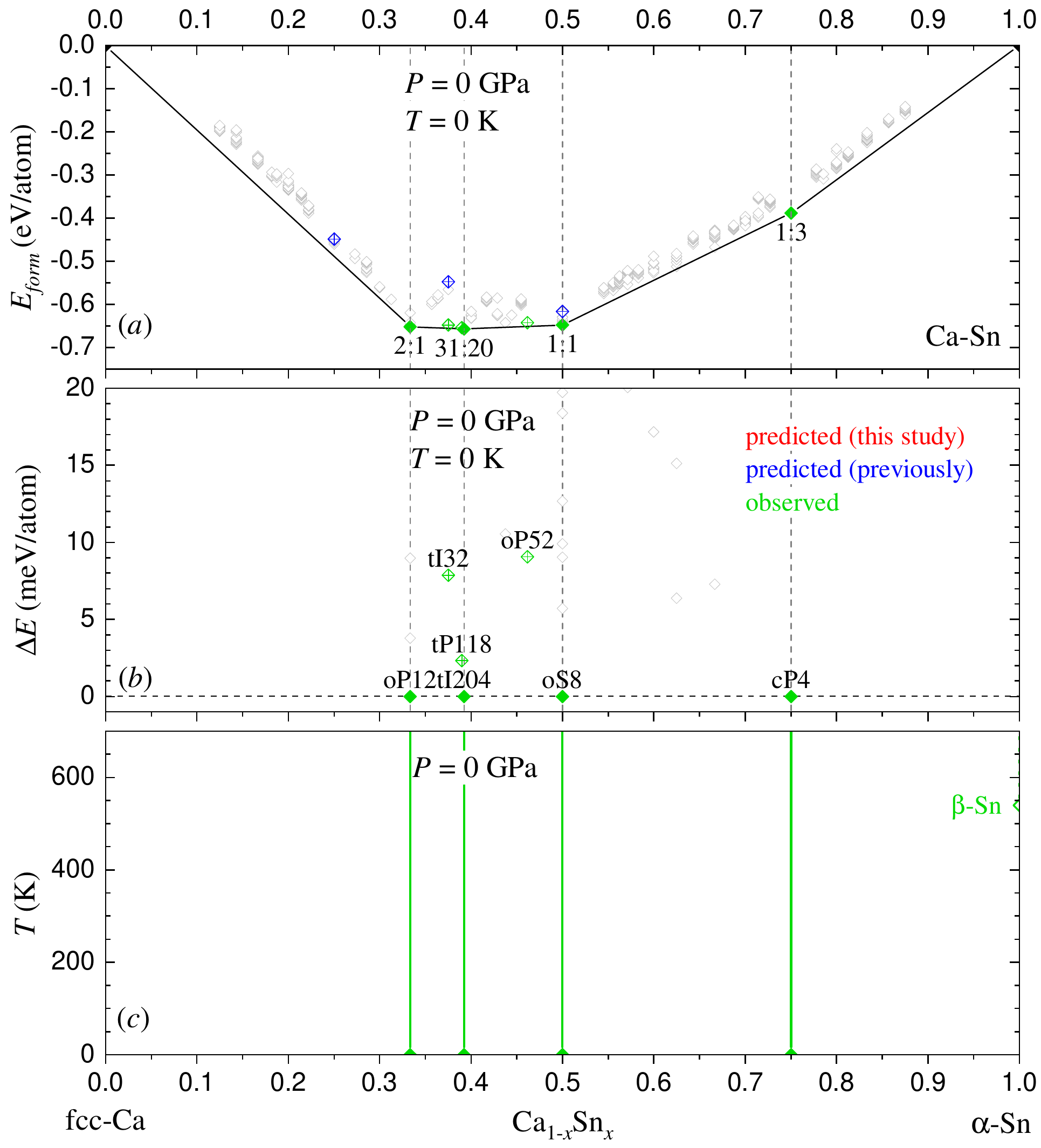} 
  \caption{Calculated stability of Ca-Sn intermetallics at ambient pressure. The three panels show (a) the convex hull constructed at $T=0$ K, (b) energy distances to the tie-line at $T=0$ K, and (c) temperature ranges of thermodynamic stability. The symbol and line styles are the same as in Fig.~\ref{fig:NaSnhull00}.}
\label{fig:CaSnhull00} 
\end{figure}

\begin{figure}[t] \centering
  \includegraphics[width=0.48\textwidth]{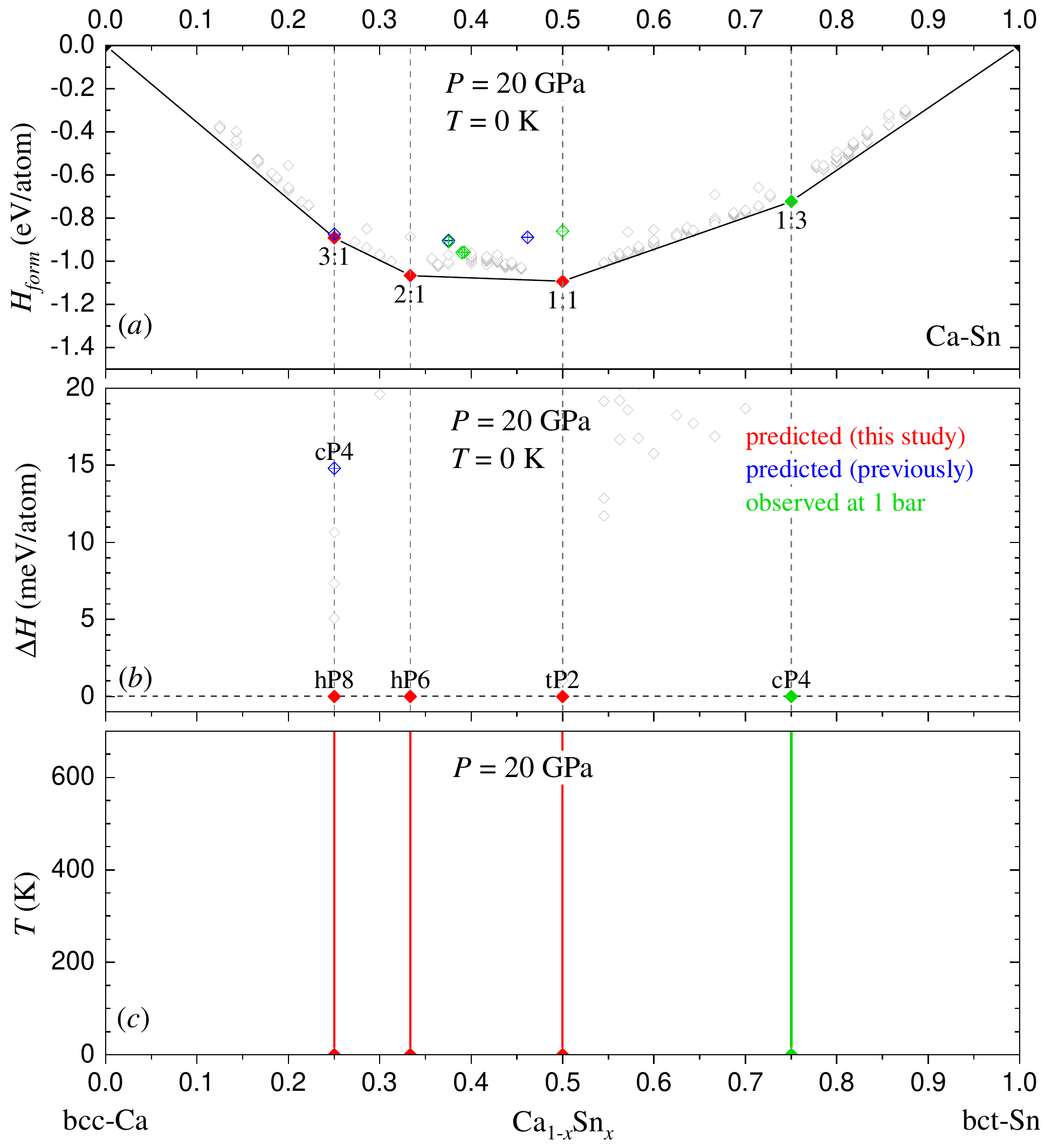} 
  \caption{Calculated stability of Ca-Sn intermetallics at 20 GPa. The three panels show (a) the convex hull constructed at $T=0$ K, (b) energy distances to the tie-line at $T=0$ K, and (c) temperature ranges of thermodynamic stability. The symbol and line styles are the same as in Fig.~\ref{fig:NaSnhull00}.} 
\label{fig:CaSnhull20} 
\end{figure}

Figs.~\ref{fig:CaSnhull00} and~\ref{fig:CaSnhull20} summarize the outcome of our Ca-Sn screening under different $(T,P)$ conditions. In line with our modeling of the other binaries, we relied on prior experimental data to include phases with more than 24 atoms per primitive cell but reproduced all other known binary intermetallics with our evolutionary structure searches. In agreement with previous DFT findings~\cite{OHNO2006,YANG2010,Yao2019}, the ambient-pressure convex hull in our calculations is defined by four oP12-Ca$_2$Sn, tI204-Ca$_{31}$Sn$_{20}$, oS8-CaSn, and cP4-CaSn$_3$ phases, while the remaining three, oP52-Ca$_7$Sn$_6$, tP118-Ca$_{36}$Sn$_{23}$, and tI32-Ca$_{5}$Sn$_{3}$, are metastable by less than 10 meV/atom. 

The compositional and formation energy proximity of the tI32-Ca$_{5}$Sn$_{3}$, tI204-Ca$_{31}$Sn$_{20}$, and tP118-Ca$_{36}$Sn$_{23}$ phases arises from their morphological connection discussed in previous studies~\cite{PALENZONA2000,Engelkemier2013}. The last two are members of the R$_{5n+6}$(T,M)$_{3n+5}$ family with $n=5$ and $n=6$ representing intergrown segments of different lengths from the parent W$_5$Cr$_3$ prototype (see Fig.~\ref{fig:CaSn-pic}(a)). In addition to being metastable, the 36:23 phase was determined previously to be mechanically unstable with $C_{44}=-122.7$ GPa~\cite{YANG2010}. The existence of the stoichiometric tI32-Ca$_{5}$Sn$_{3}$ phase itself seems to not have been established conclusively, and its appearance has been attributed to hydrogen-impurity and/or entropy-driven stabilization~\cite{PALENZONA2000,OHNO2006,Engelkemier2013}. 

\begin{figure}[!b] \centering
  \includegraphics[width=0.48\textwidth]{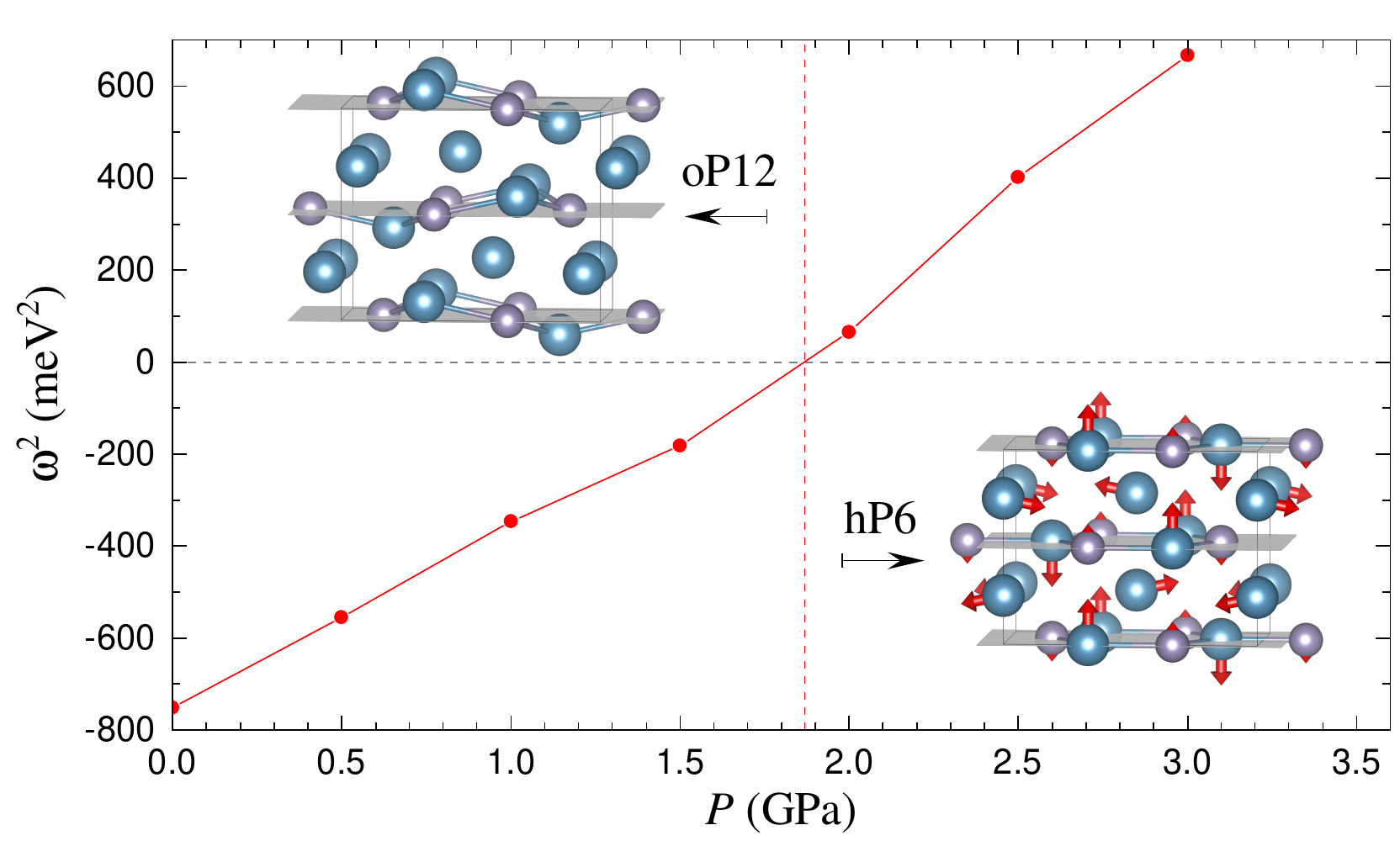} 
  \caption{Calculated $\omega^2$ dependence on pressure for a soft phonon mode defining the hP6 to oP12 transformation. The red arrows in the orthorhombic representation of the hP6 structure illustrate the atomic displacements along the eigenvector.} 
\label{fig:CaSn-extra} 
\end{figure}

Our phonon calculations indicate that the three related phases are dynamically stable and that the vibrational entropy corrections stabilize the 36:23 alloy above 1010 K but change the free energy distance of the 5:3 alloy from 7.8 meV/atom at 0 K to 5.4 meV/atom above the convex hull at 1400 K. All new Ca-Sn phases based on $\alpha$-Sn, $\beta$-Sn, and fcc-Ca unit cells proposed in Ref.~\cite{Woodcox2021} are at least 22 meV/atom above the convex hull boundary ({\it e.g.}, the blue points at $x=0.25$, 0.375, and 0.5 in Fig.~\ref{fig:CaSnhull00}).

Only one of the ambient-pressure phases, cP4-CaSn$_3$, remains thermodynamically stable at 20 GPa, which offers an opportunity to better understand and tune its intriguing topological and superconducting properties via compression. The new set of ground states identified with our global searches consists of hP8-Ca$_3$Sn, hP6-Ca$_2$Sn, and tP2-CaSn phases with simple Ni$_3$Sn, Ni$_2$In, and $\delta$-CuTi prototypes (see Fig.~\ref{fig:CaSn-pic} (c,d)). At the 2:1 composition, the known hP6 structure comprised of closely spaced heteronuclear Ca-Sn honeycomb layers intercalated with Ca is isoelectronic and morphologically related to the C32 structure adopted by MgB$_2$ but not expected to exhibit the iconic quasi-2D superconductivity defined by hole-doped B states and hard B phonon modes. The high-pressure hP6 polymorph does help appreciate the morphology of its ambient-pressure oP12 counterpart known to be a derivative of the honeycomb lattice~\cite{Lidin1998, AUDEBRAND2003}. Fig.~\ref{fig:CaSn-extra} illustrates that the frequency of the phonon mode transforming hP6 to oP12 obtained in our linear response calculations becomes imaginary below 1.9 GPa, while the linear slope of the squared frequency dependence on pressure points to a common soft-mode phase transition described by the Landau theory~\cite{Krumhansl1992,ak23}. The considerable distortions reaching 0.7 \AA\ for out-of-plane Ca displacements at 0 GPa dramatically change the local atomic environments, {\it e.g.}, changing the number of Ca neighbors around Sn atoms from 3 in hP6 to 7 in oP12 within the 3.5-\AA\ radius cutoff and from 11 in hP6 to 9 in oP12 within the 3.7-\AA\ radius cutoff. The eigenmode displacements in hP6 and the remnant honeycomb connections in oP12 displayed in Fig.~\ref{fig:CaSn-extra} help visualize the pressure-induced structural changes in Ca$_2$Sn.

%
%

\subsection{The Cu-Sn binary}

In 1990, Saunders \textit{et al.} published a detailed review of the Cu-Sn phase diagram~\cite{Saunders1990} with all (meta)stable phases discovered since the beginning of the 20th century: $\beta$~(cI2-Cu$_{17}$Sn$_{3}$), $\delta$~(cF416-Cu$_{41}$Sn$_{11}$), $\gamma$~(cF16-Cu$_{3}$Sn), $\zeta$~(hP26-Cu$_{10}$Sn$_{3}$), $\epsilon$ (oS80-Cu$_{3}$Sn), $\eta$~(hP4-Cu$_{6}$Sn$_{5}$), and $\eta^\prime$~(mS44-Cu$_{6}$Sn$_{5}$). In 1995, Larsson \textit{et al.}~\cite{Larsson1995} indexed a phase appearing above 350$^\circ$C as a monoclinic $\eta^6$~(mS54-Cu$_{5}$Sn$_{4}$). In 2009, a hP8-Cu$_{3}$Sn (D0$_{19}$) phase, labeled here as $\epsilon^\prime$, was observed by Sang \textit{et al.}~\cite{SANG2009}, which was much simpler than the previously synthesized Cu$_{3}$Ti type eight-~\cite{BERNAL1928} and ten-fold~\cite{Watanabe1983} hcp-based superstructures. 
In 2012, Wu \textit{et al.}~\cite{WU2012} characterized a new $\eta^{4+1}$ phase consisting of four $\eta^8$~(mP36-Cu$_{5}$Sn$_{4}$)~\cite{Larsson1995} and one $\eta^\prime$ unit cells. Other notable investigations of the binary system with an updated phase diagram were reported a year later~\cite{FURTAUER2013, LI2013}. In 2014, Müller and Lidin~\cite{Muller2014} performed a comprehensive characterization of Cu$_3$Sn and proposed an off-stoichiometry modulated structure to explain missing reflections in the XRD data. In our recent work~\cite{ak46}, we analyzed the thermodynamic stability of the previously reported phases at and near the 3:1 composition and found hP8 to have the lowest energy at $T=0$ K. In 2020, Leineweber \textit{et al.}~\cite{Leineweber2020} reported an experimental study of the ordered $\eta^{\prime\prime}$-Cu$_{1.235}$Sn phase along with a detailed analysis of all previously reported NiAs-type derivatives. The most recent comprehensive review of the Cu-Sn system was published in 2023~\cite{Leineweber2023} with updated phase diagrams.

The binary system has found several technological applications due to the excellent mechanical and electronic properties of Cu-Sn intermetallics. The high melting temperature, electrical conductivity, resistance to electromigration, and other beneficial characteristics have made the Cu$_{3}$Sn and Cu$_{6}$Sn$_{5}$ compounds prime candidates for Pb-free interconnects in high-performance electronic devices~\cite{yang1994,takemoto1997, Lee2001, Flandorfer2007, QU2020, BharathKrupaTeja2022, YANG2022,ak46}. The Cu-Sn alloys have also been investigated as negative electrode candidate materials in lithium-ion batteries~\cite{Kepler1999a, Kepler1999b, Wang2010, LIU2019, Tan2020}.

\begin{figure}[t] \centering
  \includegraphics[width=0.476\textwidth]{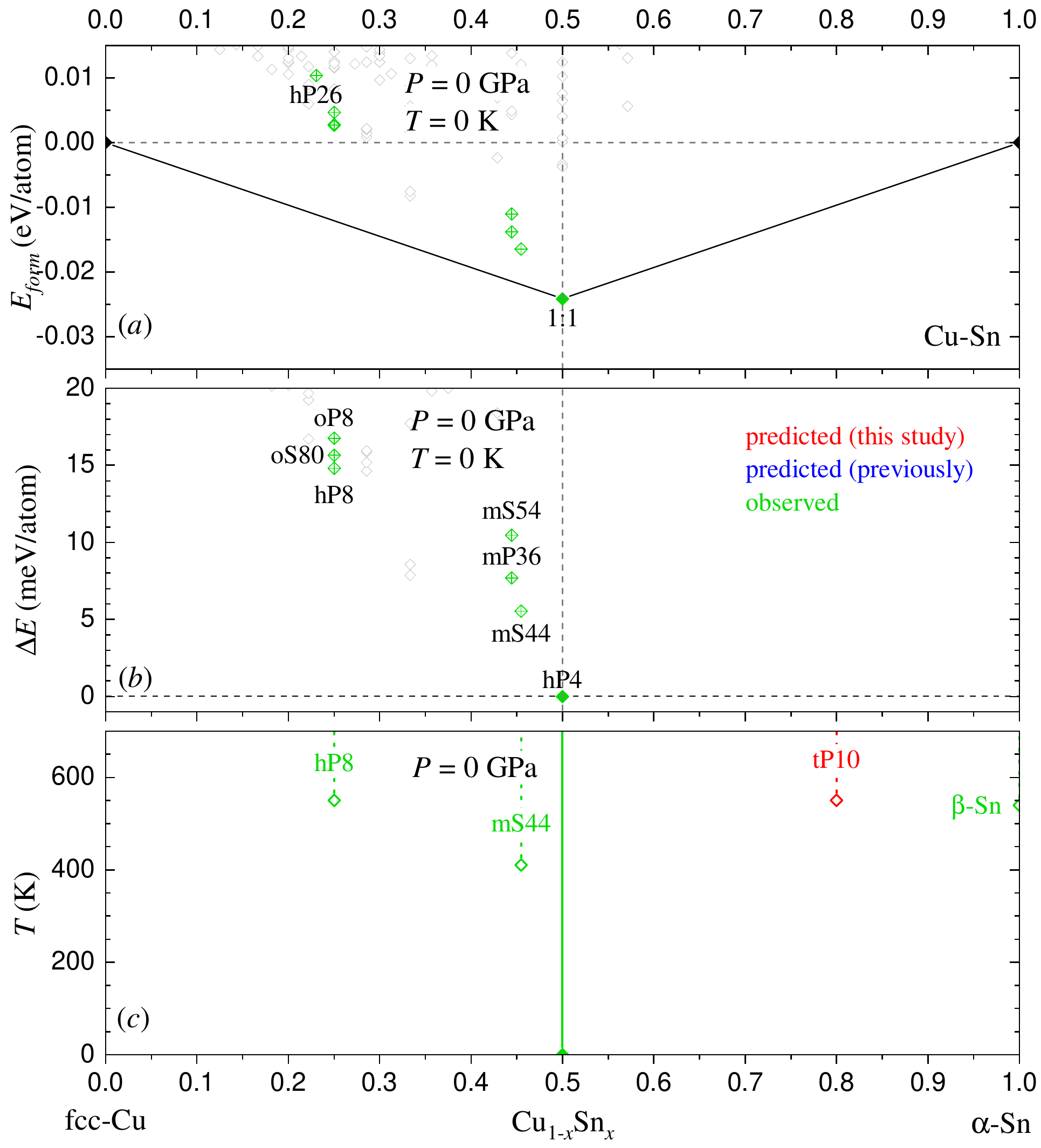} 
  \caption{Calculated stability of Cu-Sn intermetallics at ambient pressure. The three panels show (a) the convex hull constructed at $T=0$ K, (b) energy distances to the tie-line at $T=0$ K, and (c) temperature ranges of thermodynamic stability. The symbol and line styles are the same as in Fig.~\ref{fig:NaSnhull00}.}
\label{fig:CuSnhull00} 
\end{figure}

Our constructed convex hull in Fig.~\ref{fig:CuSnhull00} agrees with the current results in the Materials Project and AFLOW databases that the ordered hP4-CuSn is the only ground state with $E_{\textrm{form}} = -24$ meV/atom under ambient pressure and zero temperature in this binary system. All other observed alloys are located at least 5 meV/atom above the fcc-Cu$\leftrightarrow$hP4-CuSn tie-line, and only a handful of phases have negative formation energies. Given that only two Cu-Sn compounds, around the 6:5 and 3:1 compositions, have been reported to have stability regions extending down to the lowest displayed temperature of 373 K~\cite{Saunders1990,FURTAUER2013}, it is evident that entropy plays an important role in stabilizing the known Cu-rich alloys. The presence of the disordered $\beta$ and $\eta$ phases and the high sensitivity of transition temperature estimates to the DFT approximations~\cite{ak31,ak35} make it difficult to accurately map out the phase diagram. The following analysis focuses on assessing the importance of different contributions for the relative phase stability in this complex binary.

\begin{figure}[t] \centering
  \includegraphics[width=0.48\textwidth]{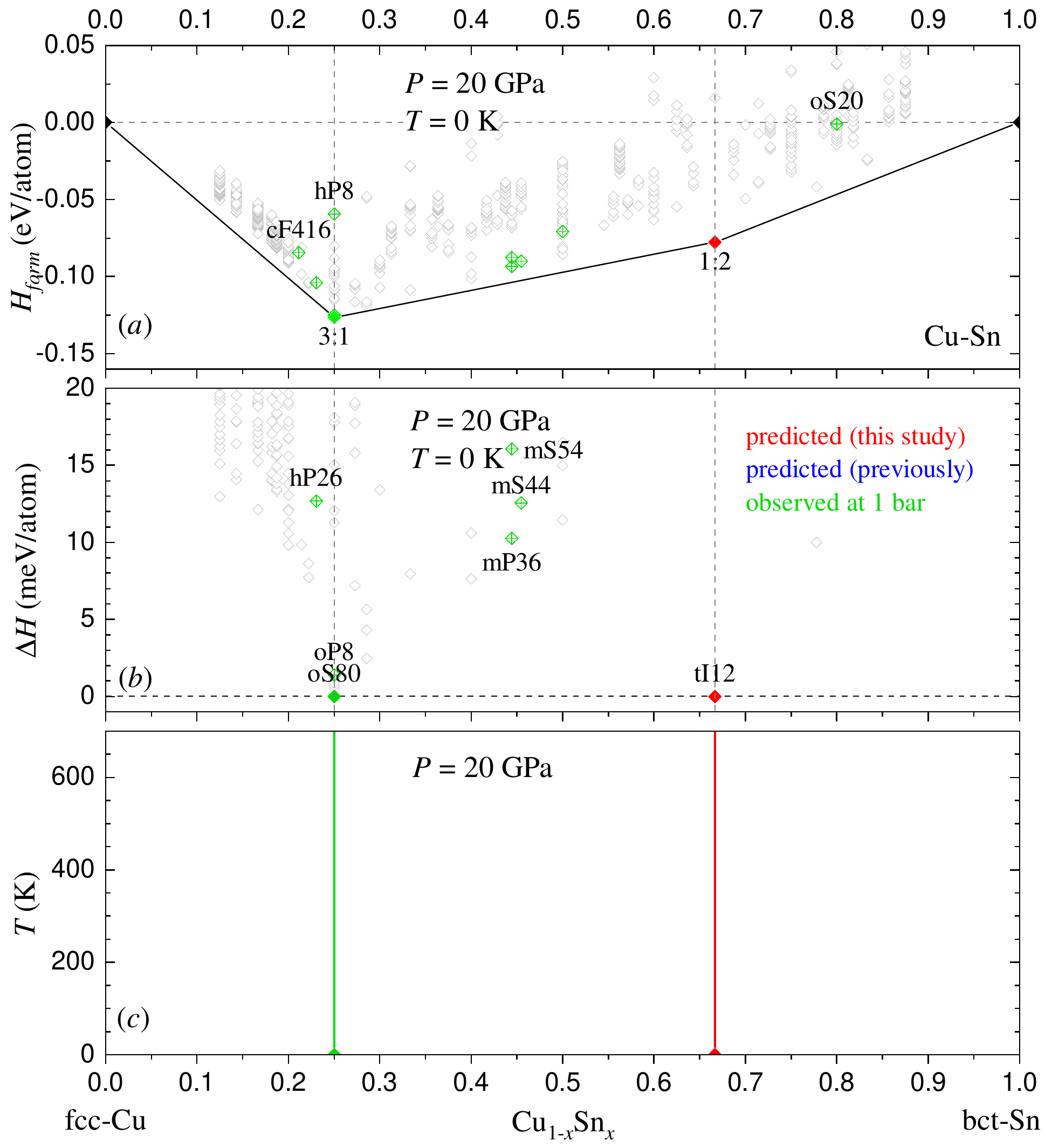} 
  \caption{Calculated stability of Cu-Sn intermetallics at 20 GPa. The three panels show (a) the convex hull constructed at $T=0$ K, (b) energy distances to the tie-line at $T=0$ K, and (c) temperature ranges of thermodynamic stability. The symbol and line styles are the same as in Fig.~\ref{fig:NaSnhull00}.} 
\label{fig:CuSnhull20} 
\end{figure}

\begin{figure}[t] \centering
  \includegraphics[width=0.48\textwidth]{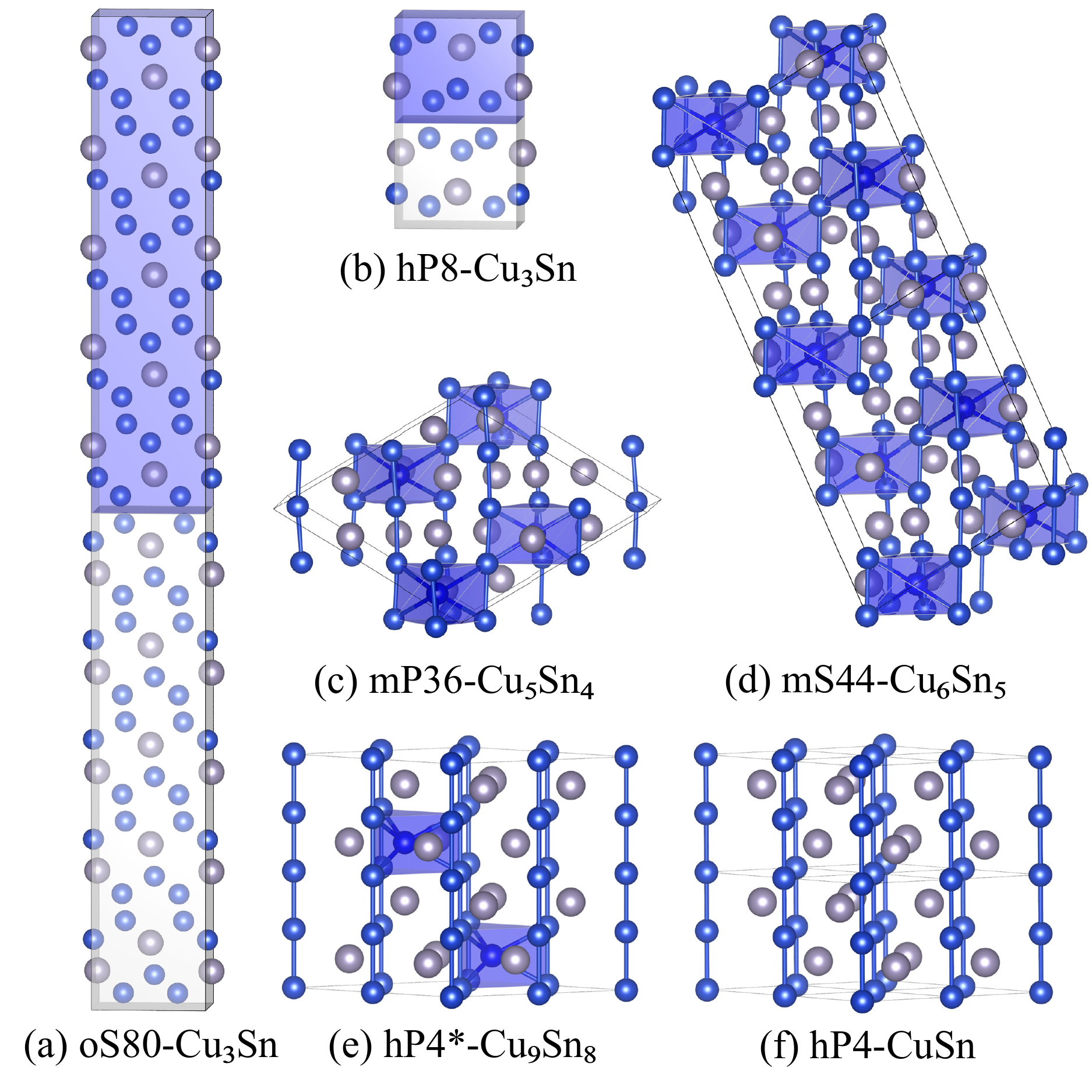} 
  \caption{ Structures of known and simulated Cu-Sn phases. (a,b) Observed hcp-based Cu$_3$Sn polymorphs, with anti-phase boundaries between blocks shaded in blue and grey. hP8 is shown in an orthorhombic representation to illustrate its relation to oS80. (c,d) Monoclinic derivatives of hP4-CuSn with ordered populations of interstitial sites. (e,f) Representation of ordered (hP4) and disordered (hP4*) phases.} 
\label{fig:CuSn-pic} 
\end{figure}

We find that the vibrational contributions indeed lower the formation free energies of all examined Cu-Sn binary phases. The results in Fig.~\ref{fig:CuSnhull00} indicate that cF416-Cu$_{41}$Sn$_{11}$ and hP26-Cu$_{10}$Sn$_{3}$, metastable by 25 meV/atom and 22 meV/atom at $T=0$ K without zero point energy, move more than half-way toward the convex hull at 550 K but do not become stable at the respective experimentally established $\sim 623$ K and $\sim 860$ K transition temperatures~\cite{FURTAUER2013}. The cF16-Cu$_{3}$Sn phase undergoes only a minor stabilization, from 75 meV/atom above the convex hull at 0 K down to 67 meV/atom at 600 K, and is certainly metastable under typical synthesis conditions. The best phase at this composition, hP8 shown in Fig.~\ref{fig:CuSn-pic}(b), has a positive formation energy of 3.6 meV/atom at 0 K but becomes a true ground state at 550 K. As discussed in our recent study~\cite{ak46}, the oS64 or oS80 superstructures with long-period anti-phase boundaries (APBs) remain a few meV/atom above hP8 at all considered temperatures and their formation is likely defined by kinetic factors. Our searches did not produce any particularly favorable bcc-based phases around $x=0.15$ to explain the stabilization of the $\beta$ alloy with entropic terms. At the Sn-rich side, the high-$T$ screening yielded a possible tP10-CuSn$_{4}$ ground state stabilizing near the reported $\eta\leftrightarrow$liquid boundary at 500 K~\cite{FURTAUER2013}. It belongs to a family of layered structures featuring metal-intercalated Sn building blocks and has the A$^+\square^-$ stacking in our notation described in the Pd-Sn subsection.

\begin{figure}[t] \centering
  \includegraphics[width=0.48\textwidth]{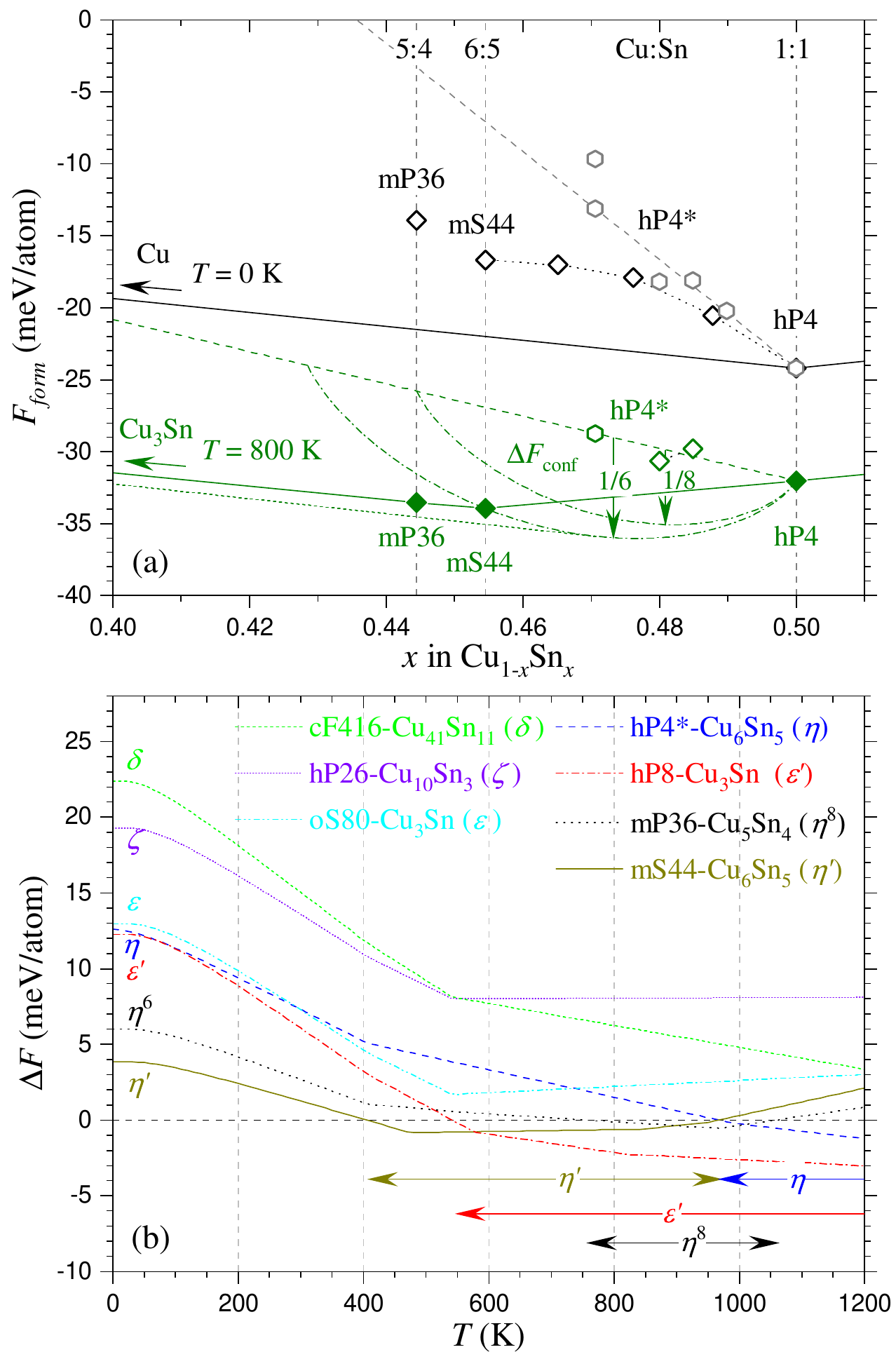} 
  \caption{ Stability analysis of Cu-Sn phases. (a) Formation free energies at 0 K (in black) and 800 K (in olive). The solid lines mark the boundaries of the convex hulls; the dashed lines are linear fits to data sets representing hP4* disordered phases with interstitial Cu defects; the dotted line connects particular interstitial configurations leading to the known mS44 monoclinic phase; the dash-dotted lines show the configurational entropy contribution with different fractions $f$ of available interstitial sites in hP4* at 800 K; and the short dash line is the tangent to the free energy curve for $f=1/6$. (b) Free energy distances to the convex hull for select Cu-Sn phases with the arrows illustrating the estimated temperature ranges of stability. Due the difficulty of modeling the disordered hP4* phase, the free energy and estimated transition temperatures were evaluated for a fixed-composition hP4*-Cu$_6$Sn$_5$ phase with $f=1/7$ occupation factor explained in the main text.}
\label{fig:CuSn-extra} 
\end{figure}

The known phases just below the 1:1 composition have a common morphology of the NiAs structure with  Cu linear chains arranged on a triangular lattice shown in Fig.~\ref{fig:CuSn-pic}(c-f). The mP36-Cu$_{5}$Sn$_{4}$, mS44-Cu$_{6}$Sn$_{5}$, and hP4*-Cu$_{6}$Sn$_{5}$ derivatives of the never-observed stoichiometric CuSn have different distributions of additional Cu atoms in trigonal bipyramidal interstices. The inclusion of the vibrational entropy does stabilize the ordered monoclinic $\eta^\prime$ above 410 K and $\eta^8$ above 760 K as illustrated in Fig.~\ref{fig:CuSn-extra}. The high-$T$ boundaries for these phases depend on the free energy of the disordered $\eta$ that benefits from an additional configurational entropy term. Its evaluation proved to be a challenge because of the sizable interaction between defects. Indeed, every CuSn unit cell with four atoms can accommodate up to two interstitial Cu atoms, which means that the compositional shifts down to $x\sim 0.45$ correspond to the trigonal bipyramidal site occupancies of $\sim 30$\%. The results in Fig.~\ref{fig:CuSn-extra}(a) for $T=0$ K demonstrate that at this occupancy level the average defect formation free energies, $\Delta F_{\textrm{d}}$, spread over a significant $0.096 - 0.213$ eV/defect range. Seeing that the most stable monoclinic decorations have fairly uniform defect distributions (Fig.~\ref{fig:CuSn-extra}), it is apparent that only a fraction of possible configurations, $f$, effectively contribute to the partition function. Moreover, the values change dramatically if the vibrational entropy contributions are included: for three representative configurations, $\Delta F_{\textrm{d}}$ was determined to decrease from 0.158, 0.154, and 0.111 eV/defect at 0 K down to 0.042, 0.024, and 0.003 eV/defect at 800 K, respectively. In light of these findings, we chose to approximate the hP4* free energy correction as $F_{\textrm{conf}}= \Delta F_{\textrm{d}} (1-2x) + k_{\textrm{B}}\frac{f}{(1+fx_d)}\{x_d\ln{(x_d)}+(1-x_d)\ln{(1-x_d)}\}$, where $x_d = \frac{(1 - 2 x)}{2fx}$, $\Delta F_{\textrm{d}} \approx 0.024 $ eV/defect, and $f$ is treated as an adjustable parameter to check correspondence with experiment. The sample curve and corresponding tangent constructed for $f=1/6$ at 800 K illustrate that hP4* can indeed easily destabilize both monoclinic phases at elevated temperatures. While this analysis does not allow us to make accurate estimates of transition temperatures, it helps rationalize the kinetics of the ordered $\eta^\prime$ and $\eta^8$ phase formation from the disordered $\eta$ precursor. Namely, it appears likely that only selected trigonal bipyramidal sites are effectively available for interstitial Cu occupation and the well-spaced defects do not need to migrate far to precipitate in the favorable ordered monoclinic configurations. The excess Cu atoms apparently do not diffuse out of the lattice at low temperatures, which may explain the absence of the stoichiometric hP4. It is worth noting that Leineweber pointed out a large variance of $E(\eta)-E(\eta’)$ values reported in several DFT studies (from -94 meV/atom to 360 meV/atom) and their discrepancy with the measured values (up to 4.5 meV/atom)~\cite{Leineweber2023}. Our proposed approach of estimating the $\eta-\eta’$ transition temperature highlights the importance of including the entropic terms in the evaluation of the free energy difference.

The application of pressure reduces the number of viable ground states down to two at the 3:1 and 1:2 compositions as shown in Fig.~\ref{fig:CuSnhull20}. It is interesting to see that the oS80-Cu$_3$Sn phase commonly obtained in ambient-pressure experiments but only metastable in DFT calculations~\cite{ak46} does stabilize over the hP8 polymorph above 10 GPa. Compression is the only factor identified in our studies so far that promotes the formation of the specific long-period superstructure. At the Sn-rich end, our NN-guided searches identified a thermodynamically stable tI12-CuSn$_{2}$ phase. This layered structure type is discussed further in the Pd-Sn and Ag-Sn subsections.

%
%

\subsection{The Pd-Sn binary}

\begin{figure}[!t] \centering
  \includegraphics[width=0.476\textwidth]{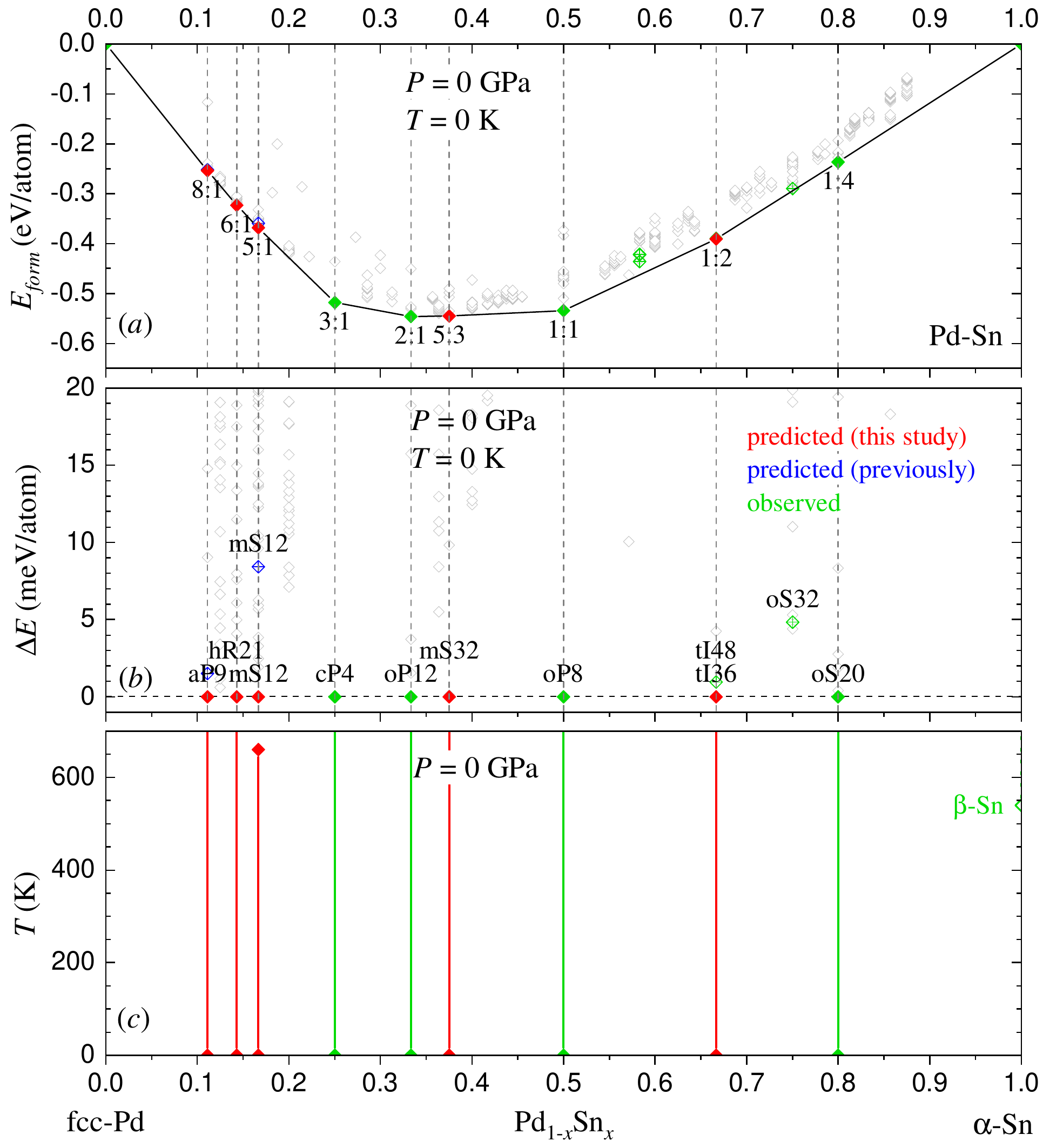} 
  \caption{Calculated stability of Pd-Sn intermetallics at ambient pressure. The three panels show (a) the convex hull constructed at $T=0$ K, (b) energy distances to the tie-line at $T=0$ K, and (c) temperature ranges of thermodynamic stability. The symbol and line styles are the same as in Fig.~\ref{fig:NaSnhull00}.}
\label{fig:PdSnhull00} 
\end{figure}

The Pd-Sn binary has been attractive primarily due to its applications in nanocatalysis~\cite{PdSn-Modibedi2011, PdSn-Geraldes2015, PdSn-Bai2017, PdSn-Li2018}. In the bulk crystalline form, Pd-Sn alloys have been observed at the 3:1, 2:1, 1:1, 5:7, 1:2, 1:3, and 1:4 compositions~\cite{Schubert1946-PdSn,Schubert1957-Pd3Sn-Pd2Sn,Romaka2010-Pd5Sn7,Hellner-PdSn2,Schubert1958-PdSn3,PdSn-Nylen2004}. The oS20-PdSn$_4$ phase with space group $Ccca$ has received most attention due its unusual topological properties. This Sn-rich phase has the same crystal structure and electron count as PtSn$_4$~\cite{PdSn-Kubiak1984, PdSn-Nylen2004}, known for its extremely large magnetoresistance~\cite{PdSn-Mun2012} and Dirac node arcs~\cite{PdSn-Wu2016, PdSn-Jo2017}, but features gapped out Dirac node arcs~\cite{PdSn-Jo2017}.

\begin{figure}[!t] \centering
  \includegraphics[width=0.48\textwidth]{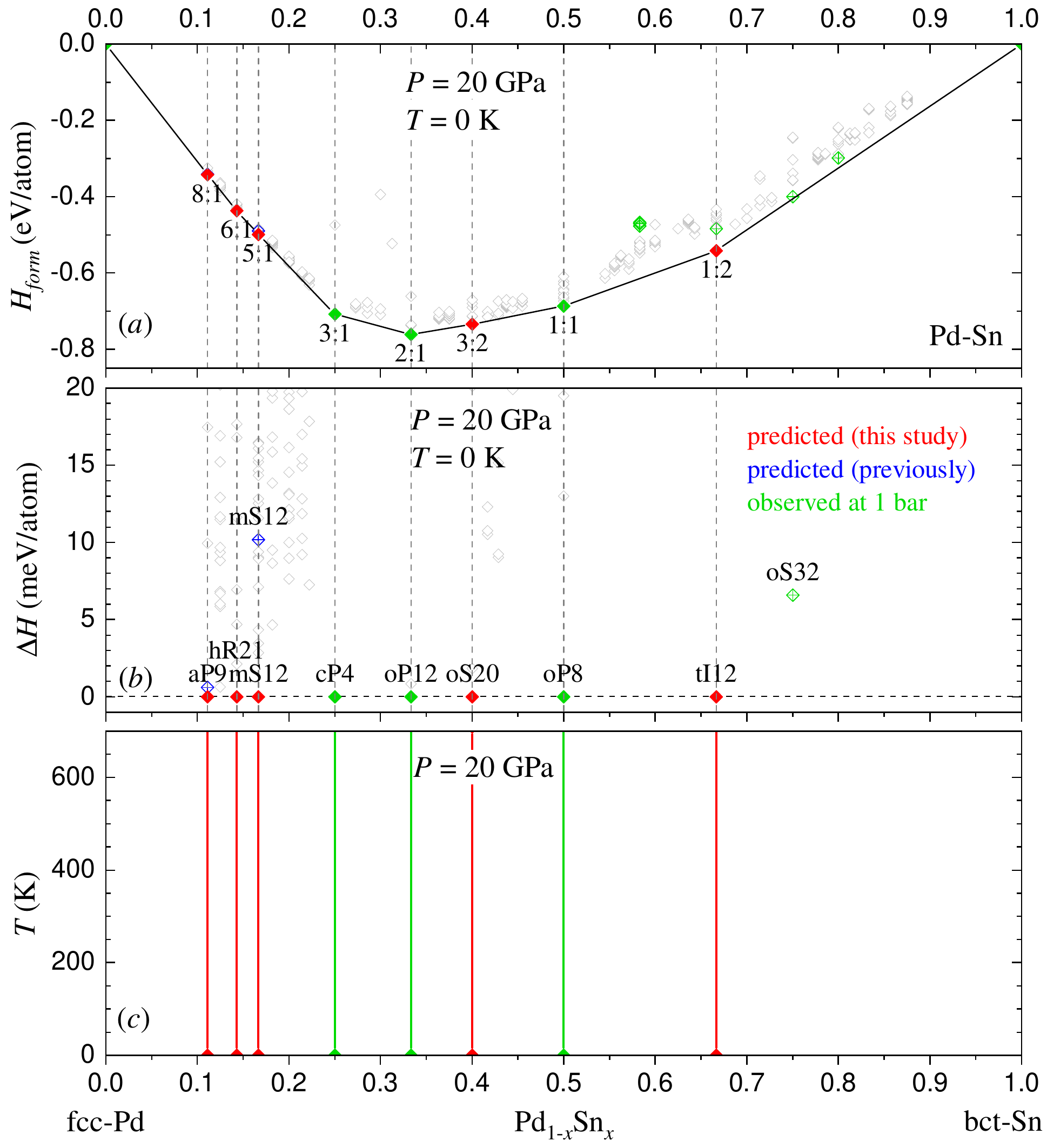} 
  \caption{Calculated stability of Pd-Sn intermetallics at 20 GPa. The three panels show (a) the convex hull constructed at $T=0$ K, (b) energy distances to the tie-line at $T=0$ K, and (c) temperature ranges of thermodynamic stability. The symbol and line styles are the same as in Fig.~\ref{fig:NaSnhull00}.} 
\label{fig:PdSnhull20} 
\end{figure}

Our screening and modeling of Pd-Sn alloys indicate that the binary may have several additional phases synthesizable at both ambient and high pressures, as illustrated in Figs.~\ref{fig:PdSnhull00} and~\ref{fig:PdSnhull20}. The known phases defining parts of the ambient-pressure convex hull are oP8-PdSn discovered in 1946~\cite{Schubert1946-PdSn}, cP4-Pd$_3$Sn and oP12-Pd$_2$Sn reported in 1957~\cite{Schubert1957-Pd3Sn-Pd2Sn}, and oS20-PdSn$_4$ observed in 2004~\cite{PdSn-Nylen2004}. The mP24-Pd$_5$Sn$_7$ phase was shown in 2010 to have partial occupancies~\cite{Romaka2010-Pd5Sn7}. We simulated six different decorations of the 4h sites in unit cells with Sn atoms and observed that the fully relaxed structures lie 24-40 meV/atom above the tie-line. The tI48-PdSn$_2$~\cite{Hellner-PdSn2} and oS32-PdSn$_3$~\cite{Schubert1958-PdSn3} phases observed in the 1950s were found to be metastable by about 1 and 5 meV/atom, respectively.

The majority of new competitive compounds found in our searches are located at the Pd-rich end of the composition range. The tI18-Pd$_8$Sn phase matches the bf0bcb581de49d90 entry listed in AFLOW but not discussed previously. It is based on the Pd bcc lattice with Sn substitutions that lead to a $c/a=1.055$ tetragonal distortion and a $-21.3$ meV/atom stabilization with respect to the fcc-Pd$\leftrightarrow$cP4-Pd$_3$Sn tie-line. Although 8:1 was outside the range of standard compositions we chose to scan for the Sn binaries, the presence of the tI18 putative ground state prompted us to perform a search at this stoichiometry as well. Our resulting best candidate, aP9-Pd$_8$Sn, is an improvement on tI18 by 1.9 meV/atom. The identified aP8-Pd$_7$Sn, hR21-Pd$_6$Sn, and mS12-Pd$_5$Sn phases happen to be different decorations of the fcc lattice that break the fcc-Pd$\leftrightarrow$cP4-Pd$_3$Sn tie-line by similar $-24.3$, $-27.0$, and $-22.2$ meV/atom, and lie $0.6$, $-3.8$, and $-1.2$ meV/atom from the aP9-Pd$_8$Sn$\leftrightarrow$cP4-Pd$_3$Sn tie-line, respectively. The mS12-Pd$_5$Sn phase found in our searches is 8.1 meV/atom more stable than the proposed mS12-Pd$_5$Sn reported by Wang {\it et al.}~\cite{Wang2021}. The relative energies are consistent in other DFT approximations (see Table S4), which suggests that at least some of these four ground states might be obtained via standard synthesis routes. We also find that an mP32-Pd$_5$Sn$_3$ phase with isolated Sn atoms breaks the oP12-Pd$_2$Sn$\leftrightarrow$oP8-PdSn tie-line by $-1.7$ meV/atom and a tI36-PdSn$_2$ phase is below tI48-PdSn$_2$ experimental phase by $-1.3$ meV/atom. Fig.~\ref{fig:PdSnhull00}(c) shows that the vibrational contributions do not cause any changes in the convex hull up to 600 K.

\begin{figure}[!t] \centering
  \includegraphics[width=0.48\textwidth]{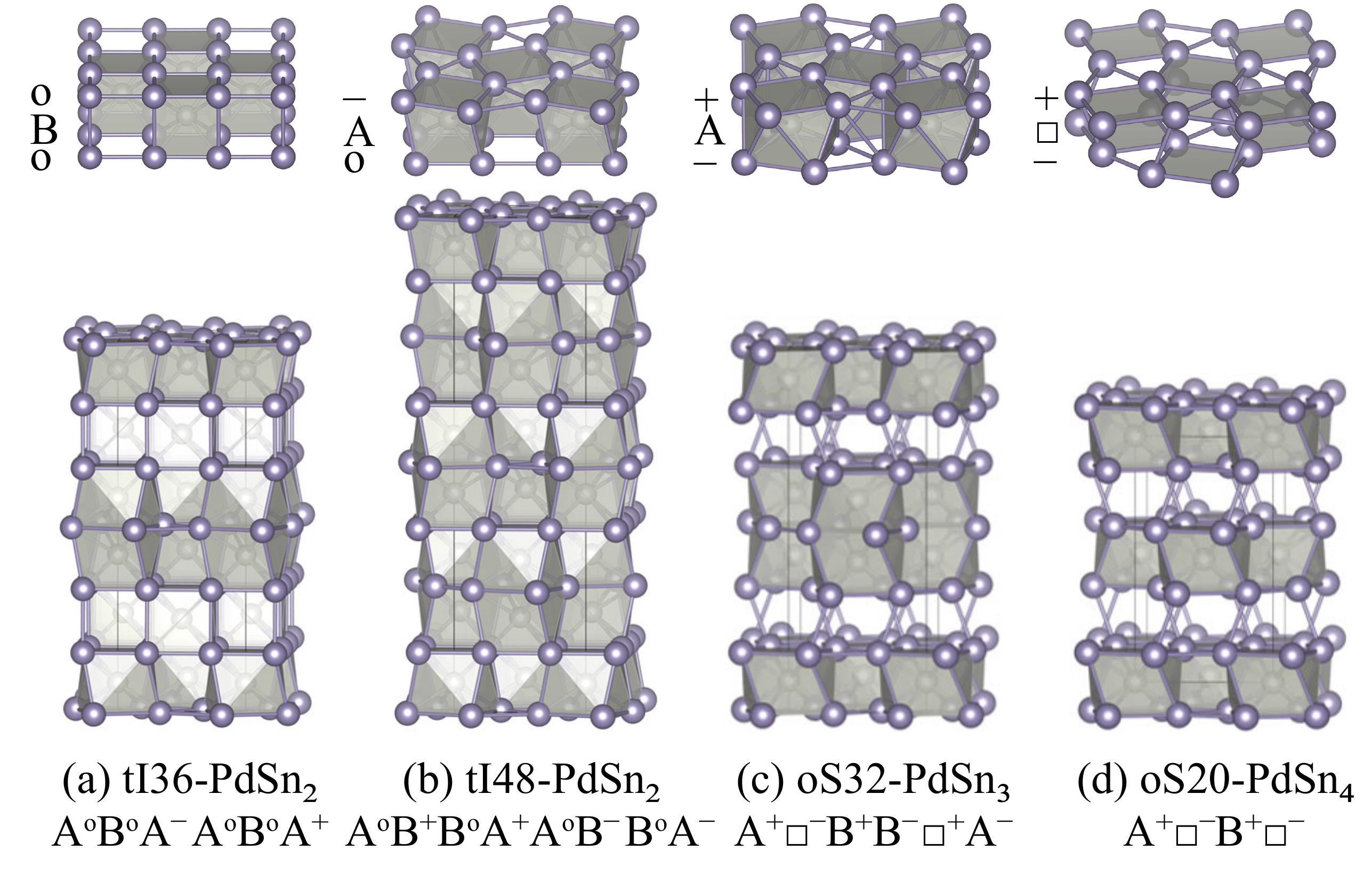} 
  \caption{Layer stacking of Pd-rich phases at ambient pressure. In this notation, A or B determines if the Pd atoms are in the (0,0) and (1/2,1/2) positions or in the (1/2,0) and (0,1/2) positions, while an empty square symbol denotes the absence of a Pd layer. For the Sn layers, the "o" superscript denotes a layer of squares (4$^4$ Kepler nets), while "+" and "$-$" superscripts specify the square rotation direction within the layer of squares and rhombi (3$^2$434 Kepler nets) as described in the text.}
\label{fig:PdSnStacking} 
\end{figure}

In order to better understand stable structural motifs defining the Sn-rich ground states, we compared the proposed and known phases with tetragonal (tI36-PdSn$_2$ and tI48-PdSn$_2$) or near-tetragonal (oS32-PdSn$_3$ and oS20-PdSn$_4$) symmetry featuring eight-fold coordinated Pd atoms (see Fig.~\ref{fig:PdSnStacking}). According to the detailed analysis of MSn$_n$ ($n=2-4$) alloys by Nylen \textit{et al.}~\cite{PdSn-Nylen2004}, observed Sn-rich phases in several binary systems can be described as sequences of closely related building blocks. The Sn layers appear as either unrotated squares (4$^4$ Kepler nets in Schl{\"a}fli notation) or a combination of rotated squares by 15-20$^{\circ}$ and rhombi (3$^2$434 Kepler nets in Schl{\"a}fli notation). Blocks formed by adjacent Sn layers along the stacking axis can also be shifted by (1/2,0), (0,1/2), or (1/2,1/2) within the basal plane.

The stacking sequences in such phases can be represented conveniently with an alternative notation that explicitly specifies the location and orientation of the Pd and Sn layers. The "A", "B", and "$\square$" symbols denote A-centered, B-centered, and missing Pd layers, respectively (see Fig.~\ref{fig:PdSnStacking}). The "$+$", "$-$", and "o" superscripts indicate clockwise, counterclockwise, and null rotations of Sn squares, respectively. The lateral placement of Sn squares is defined by the centering of the adjacent Pd layer(s) because metal atoms are never found directly above or below the Sn rhombi. The rotation sign is defined for Sn squares centered at (1/2,0) or (1/2,1/2). With this convention, the new tI36-PdSn$_2$ member of the PdSn$_n$ family can be represented as A$^\circ$B$^\circ$A$^-$A$^\circ$B$^\circ$A$^+$ (Fig.~\ref{fig:PdSnStacking}(a)), which illustrates the presence of near-cubic $^\circ$B$^\circ$ Sn coordinations around Pd atoms not seen in the known Pd-Sn prototypes.

\begin{figure}[!t] \centering
  \includegraphics[width=0.48\textwidth]{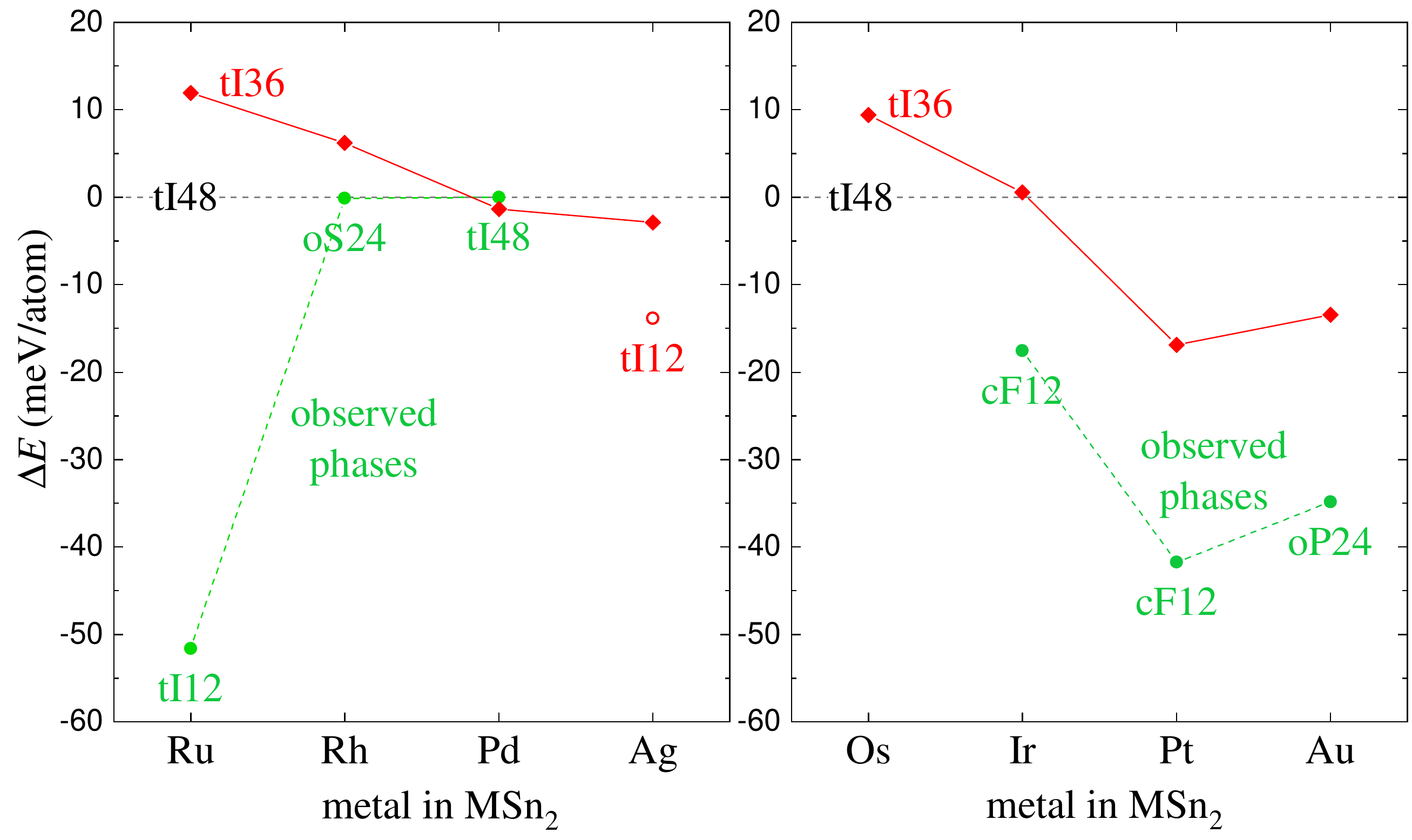} 
  \caption{Relative stability of MSn$_2$ polymorphs referenced to the known tI48 structure. The red solid diamonds correspond to the proposed tI36 structure. The green solid circles denote the most stable known phase for each existing MSn$_2$ compound. The red hollow circle marks the predicted tI12-AgSn$_2$ discussed in the main text.}
\label{fig:PdSnNoble} 
\end{figure}

Considering the ubiquity of observed MSn$_2$ compounds, it seemed fitting to examine the relative stability of the competing prototypes across the block of noble metals. We chemically substituted and fully relaxed binary Sn alloys with Ru, Rh, Pd, Ag, Os, Ir, Pt, and Au metals in the tI36 and tI48 unit cells. Our results presented in Fig.~\ref{fig:PdSnNoble} show that tI36 becomes favored over tI48 at the electron-rich end of both $4d$ and $5d$ sets. Comparison of these phases to the ground states appearing in the Materials Project database~\cite{MatProj} shows that only PdSn$_2$ has the potential to form in the tI36 configuration and that RhSn$_2$~\cite{Hellner-PdSn2} has virtually identical oS24 and tI24 energies due to a close relationship between the two prototypes with A$^\circ$B$^+$B$^\circ$A$^+$ and A$^\circ$B$^+$B$^\circ$A$^+$A$^\circ$B$^-$B$^\circ$I$^-$ representations.

The hydrostatic compression to 20 GPa induces few changes in the calculated convex hull (Fig.~\ref{fig:PdSnhull20}). Our predicted ambient-pressure mS32-Pd$_5$Sn$_3$ ground state is replaced with a new one, oS20-Pd$_3$Sn$_2$, at a nearby composition. At the 1:2 stoichiometry, the tI12-PdSn$_2$ polymorph stabilizes over tI48 by 58.4 meV/atom. This structure with the A$^+$A$^-$ sequence (see Fig.~\ref{fig:AgSnPic}) is also predicted to be favored at elevated pressure for CuSn$_2$ (Fig.~\ref{fig:CuSnhull20}) and at elevated temperature for AgSn$_2$ (Fig.~\ref{fig:AgSnhull00}). On the Sn-rich side, the known oS32-PdSn$_3$ phase remains metastable and the PdSn$_4$ compound is no longer favored.

%
%

\subsection{The Ag-Sn binary}

The most comprehensive Ag-Sn phase diagram was compiled by Karakaya and Thompson in 1987~\cite{Karakaya1987}. It summarized an extensive body of research on Ag-Sn alloys carried out since the 1920s. The accrued data indicated that the binary system has a high solubility of Sn in fcc-Ag up to $x=0.115$, a disordered $\zeta$ alloy between $x=0.118$ and $x=0.228$, an ordered Ag$_3$Sn compound $\epsilon$ with a relatively narrow stability range, and a eutectic point at the Sn-rich end with $x=0.962$ and 221°C. The $\zeta$ alloy was identified as an hcp-based solid solution with a near-ideal $c/a=1.626$. Ag$_3$Sn has also been determined to have an underlying hcp lattice but its exact crystal structure has been the subject of a long debate. Structural solutions proposed since the compound’s discovery in 1926~\cite{Murphy1926} include a disordered hcp phase~\cite{preston1926}, an orthorhombic displacive homeotype $\beta$-TiCu ($Cmcm$)~\cite{ellner2003}, and the D0$_a$-Cu$_3$Ti prototype ($Pmmn$)~\cite{fairhust1972}. The latest studies~\cite{novakovic2018,yu2018,saleh2018,zhou2020,cui2023} point to a general consensus that the equilibrium $\epsilon$ phase has the ordered Cu$_3$Ti-type structure (oP8).

\begin{figure}[t] \centering
  \includegraphics[width=0.476\textwidth]{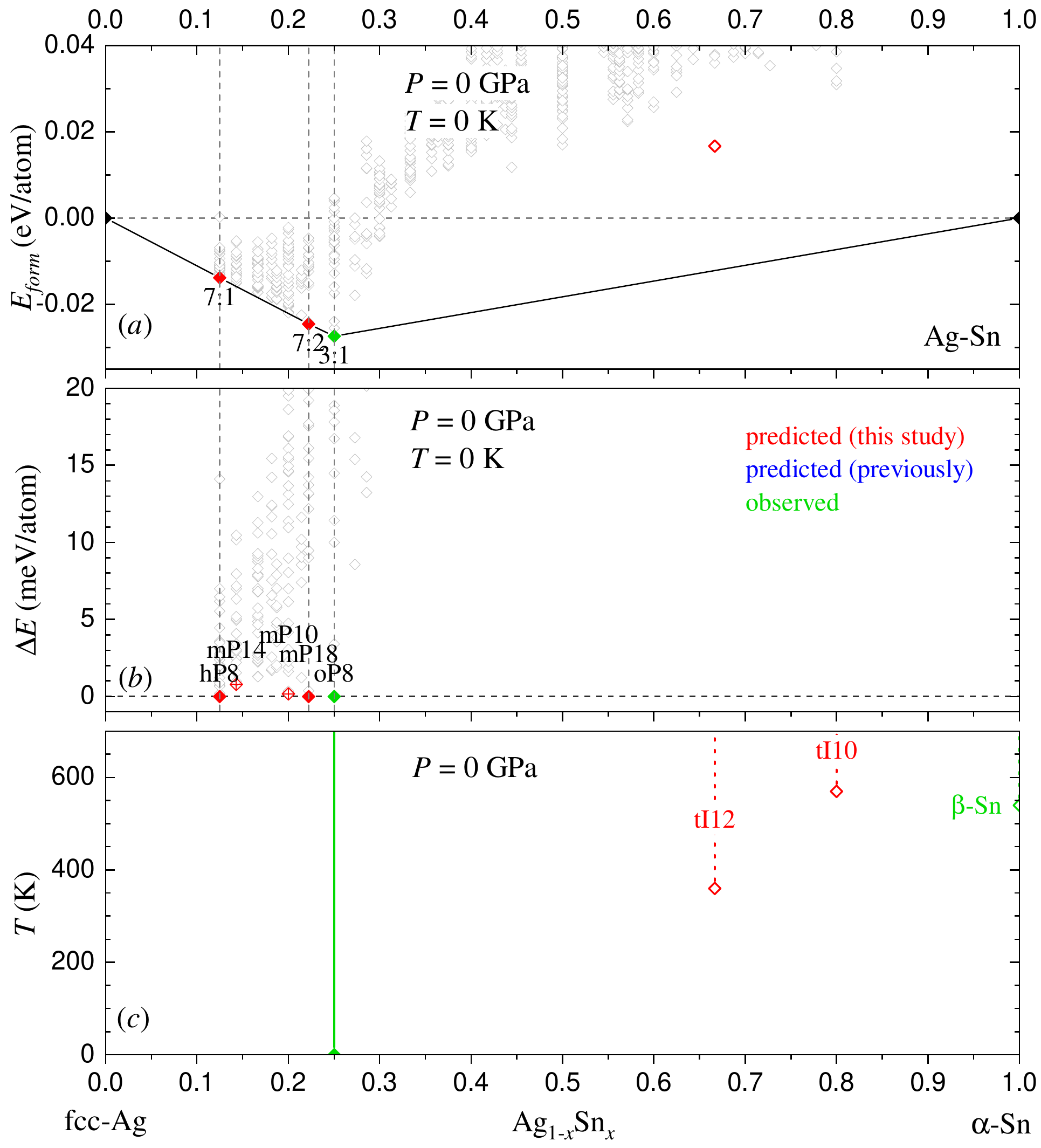} 
  \caption{Calculated stability of Ag-Sn intermetallics at ambient pressure. The three panels show (a) the convex hull constructed at $T=0$ K, (b) energy distances to the tie-line at $T=0$ K, and (c) temperature ranges of thermodynamic stability. The symbol and line styles are the same as in Fig.~\ref{fig:NaSnhull00}.}
\label{fig:AgSnhull00} 
\end{figure}

\begin{figure}[t] \centering
  \includegraphics[width=0.48\textwidth]{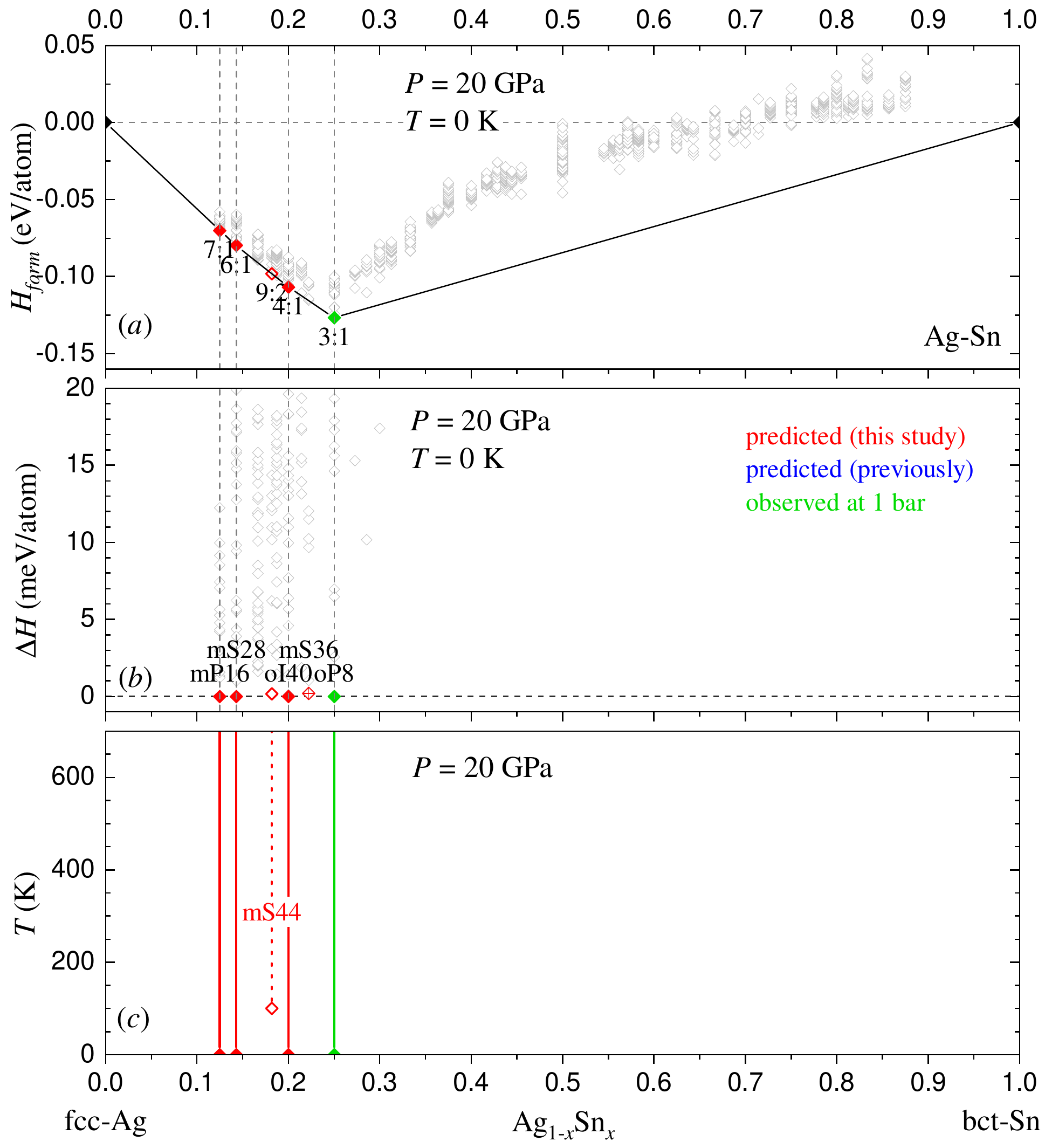} 
  \caption{Calculated stability of Ag-Sn intermetallics at 20 GPa. The three panels show (a) the convex hull constructed at $T=0$ K, (b) energy distances to the tie-line at $T=0$ K, and (c) temperature ranges of thermodynamic stability. The symbol and line styles are the same as in Fig.~\ref{fig:NaSnhull00}.} 
\label{fig:AgSnhull20} 
\end{figure}

Study of Ag-Sn compounds is of significant importance in the ongoing development of Pb-free solders for high-performance electronic devices~\cite{Moon2000,Li2005,Rossi2016,novakovic2018,cui2023}. The binary eutectic intermetallic is an integral part of the Sn-rich multicomponent alloys with Cu, Sb, Bi, and In optimized for robust operation under mechanical and thermal stresses. In particular, the extensive work discussed in Ref.~\cite{cui2023} has demonstrated that joints’ resistance to creep and thermal fatigue is strongly affected by the morphology of Ag$_3$Sn forming during solder solidification. A recent transport study has also indicated that Ag$_3$Sn is a topological material with a nontrivial Berry phase and a possible candidate for valleytronics~\cite{zhou2020}.

Our screening and modeling findings (see Fig.~\ref{fig:AgSnhull00} and \ref{fig:AgSnhull20}) are consistent with the experimental observations. The unconstrained searches identified a number of hcp-based Ag-Sn phases in the $0.12<x<0.23$ range that lie within 1-2 meV/atom to the fcc-Ag$\leftrightarrow$oP8-Ag$_3$Sn tie line. At the 7:1 composition, for instance, a single Ag substitution for Sn in the hexagonal $2\times 2\times 1$ hcp-Ag supercell ($c/a=1.638$) produced an hP8 structure ($c/a=1.620$) located essentially on the tie-line (0.2 meV/atom in PBE, -1.1 meV/atom in LDA and 0.7 meV/atom in SCAN). An alternative monoclinic decoration resulted in a nearly degenerate mP16-Ag$_7$Sn phase (0.4 meV/atom in PBE, $-1.8$ meV/atom in LDA and 1.2 meV/atom in SCAN). At the 7:2 composition, a monoclinic mP18 phase was found to be similarly close to stability (0.2 meV/atom in PBE, $-0.7$ meV/atom in LDA and 1.5 meV/atom in SCAN). It is evident that the hcp derivatives have comparable energies, which explains their appearance as a solid solution promoted by the configurational entropy in this Ag-rich part of the phase diagram. In agreement with the previous {\it ab initio} analysis of the random alloy, the most stable hcp configurations have fairly uniform distributions of Sn atoms~\cite{saleh2018}.

To better understand the favorability of oP8 at the 3:1 stoichiometry, we compared the relative stability of competing hcp supercells for the binary M$_3$Sn compounds with Cu, Ag, and Au. Fig. S11 illustrates that only Cu$_3$Sn benefits from the appearance of APBs and has a slight preference, by $2.4$ meV/atom, for the hP8 configuration with the highest APB number per formula unit over oP8. In the M$_3$Sn compounds with the larger Ag and Au metals, the trend is reversed and oP8 is favored over hP8 by 4.9 meV/atom and 19 meV/atom, respectively. We also considered the simplest orthorhombic representation of hcp at the Ag$_3$Sn composition with the reported unit cell dimensions of $a = 2.98$ \AA, $b = 5.15$ \AA, and $c = 4.77$ \AA~\cite{ellner2003}. In the ordered form, this four-atom unit cell cannot have the $Cmcm$ symmetry. We found the fully relaxed oP4 structure with the $Pmm2$ symmetry to be less stable than oP8 by 43 meV/atom.

\begin{figure}[t!] \centering
  \includegraphics[width=0.48\textwidth]{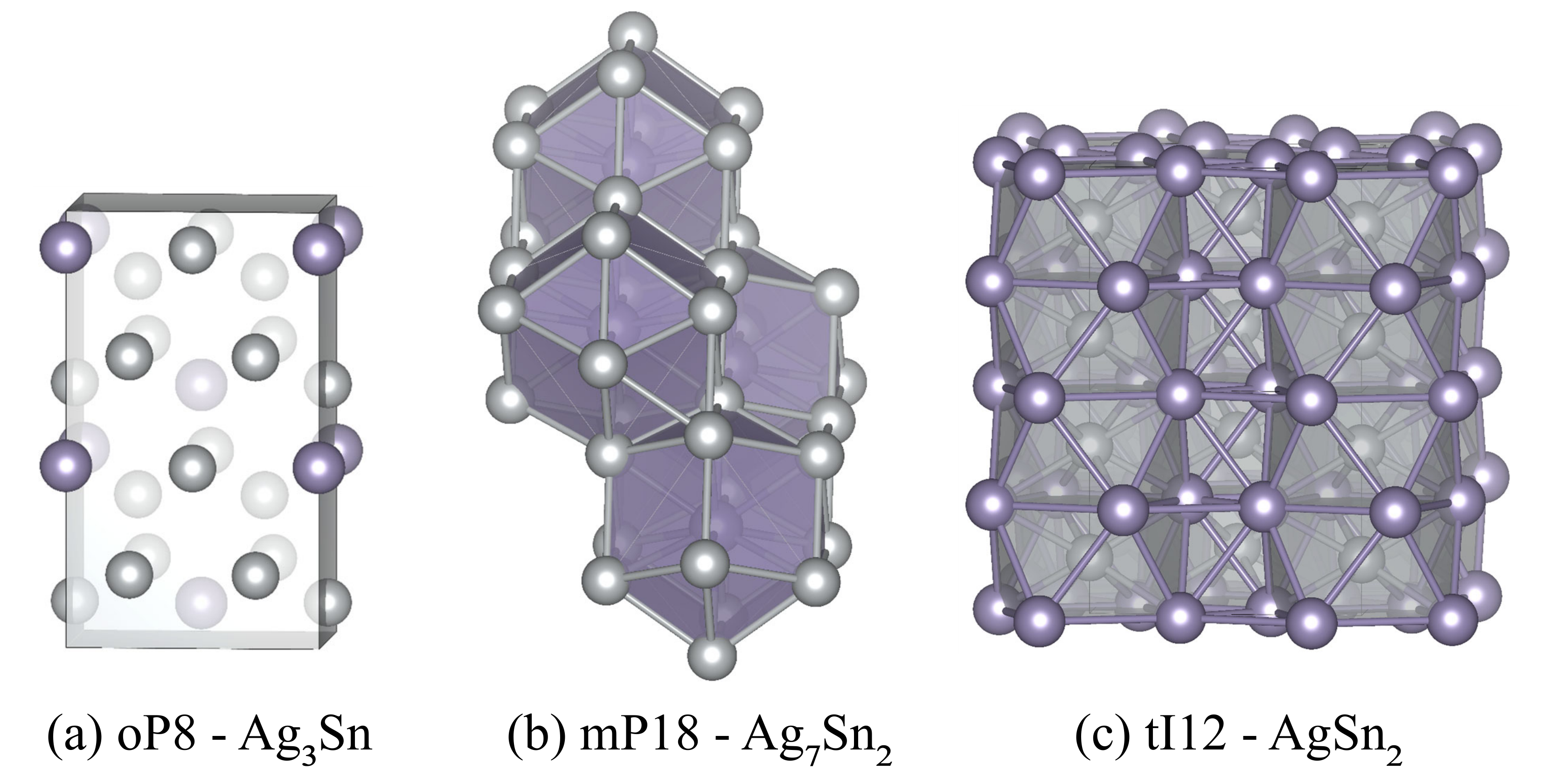} 
  \caption{ Structures of relevant Ag-Sn phases. (a) oP8-Ag$_3$Sn is the only experimentally observed phase in the Ag-Sn binary system and a subject of multiple studies to date. (b) mP18-Ag$_7$Sn$_2$ is one of the ordered representations of the disordered $\zeta$ phase. (c) A$^+$A$^-$ tI12-AgSn$_2$ is a predicted high-$T$ ground state with the stacking notation explained in the Pd-Sn subsection. } 
\label{fig:AgSnPic} 
\end{figure}

The inclusion of the vibrational entropy has little effect on the relative free energies for phases with $x \le 0.25$ but stabilizes two new phases at the Sn-rich end in our calculations. The tI12--AgSn$_2$ phase, shown to be metastable by Saleh {\it et al.}~\cite{saleh2018}, is found 29 (6) meV/atom above the oP8-Ag$_3$Sn$\leftrightarrow\alpha$-Sn ($\beta$-Sn) tie-line in our $T$=0~K calculations as well. This structure has the familiar A$^+$A$^-$ stacking (see Fig.~\ref{fig:AgSnPic}(c)) appearing as the high-$P$ ground states in the Cu-Sn and Pd-Sn binaries. In the tI12-AgSn$_2$ case, the entropic contribution makes it the ground state above 360 K. The tI10-AgSn$_4$ phase breaks the tI12-AgSn$_2\leftrightarrow\beta$-Sn tie-line at much higher 570~K temperature that falls between the $\sim 500$~K and $\sim 680$~K boundaries that define the stability range of the $\epsilon$-liquid mixture at this composition. The tall tetragonal unit cell with $a=3.310$ \AA\ and $c=23.51$ \AA\ has two Ag atoms regularly spaced along the $c$-axis comprised of five stacked bct blocks. The application of 20 GPa compression (Fig.~\ref{fig:AgSnhull20}) destabilizes the two Sn-rich phases and leads to minor reordering of relative enthalpies, by 1-2 meV/atom, for competing hcp derivatives with $0.12<x<0.25$. Some of these ordered phases, such as mP16 and oI40, become marginally stable at elevated pressures and temperatures.

\section{Summary}
\label{summary}

The NN-guided materials prediction approach introduced in our previous study and tested on a single Li-Sn binary system~\cite{ak47} has been used in this work to screen a larger configuration space of five M-Sn binaries. Over 2.0 million candidate phases were fully optimized with NN models in our evolutionary searches, requiring over 260,000 CPU hours, and almost 14,000 of them were further re-examined with DFT, costing about 0.98 million CPU hours. Compared to our previous study of Li-Sn alloys~\cite{ak47}, the NN simulations for similar-sized structures were cheaper because of the lower number of nearest neighbors within the cutoff radius for the Sn binaries with larger metals, while the DFT calculations were more expensive because of the higher number of included (semi-core) electrons. Therefore, an equivalent DFT-level scan of the M-Sn phases with the considered structure sizes would have necessitated about 100 million CPU hours, which indicates that the utilization of NN potentials accelerates global structure searches at $T=0$ K by two orders of magnitude. An even more considerable benefit was seen in the performed screening for high-$T$ ground states, as our phonon calculations at the NN level for over 6,000 phases (230 CPU hours) helped narrow the pool down to 173 viable candidates for subsequent phonon calculations at the DFT level (1.2 million CPU hours).

\begin{table}[!t] 
  \begin{tabularx}{\linewidth} {l c c r c l r c l C} \hline \hline
  Tin & Pearson & Space & \multicolumn{6}{c}{$T$ range (K)} & AFLOW  \\ 
  alloy & symbol & group & \multicolumn{3}{c}{0 GPa} & \multicolumn{3}{c}{20 GPa} & prototype \\ 
  \hline
 Na$_{4}$Sn        & mP20 & 11  & 250 & - &     &   &   &     & - \\
 NaSn$_{2}$        & hP6  & 194 & 70  & - &     &   &   &     & CaIn$_2$ \\
 NaSn$_{5}$        & oI12 & 71  & 450 & - &     &   &   &     & - \\
 Na$_{4}$Sn        & hR15 & 166 &     &   &     & 0 & - &     & - \\
 Na$_{11}$Sn$_{3}$ & hR42 & 166 &     &   &     & 0 & - &     & - \\
 Na$_{7}$Sn$_{2}$  & oS36 & 65  &     &   &     & 0 & - &     & - \\
 Na$_{3}$Sn        & hP12 & 156 &     &   &     & 0 & - & 630 & - \\
 Na$_{2}$Sn        & hP3  & 191 &     &   &     & 0 & - &     & Os$_2$Pt \\
 NaSn              & mP6  & 10  &     &   &     & 0 & - &     & LiSn \\
 Na$_{3}$Sn$_{5}$  & oS16 & 65  &     &   &     & 0 & - & 420 & - \\
 NaSn$_{3}$        & cP4  & 221 &     &   &     & 0 & - &     & Cu$_3$Au (L1$_2$) \\
 \hline
 Ca$_{3}$Sn        & hP8  & 194 &     &   &     & 0 & - &     & Ni$_3$Sn (D0$_{19}$) \\
 Ca$_{2}$Sn        & hP6  & 194 &     &   &     & 0 & - &     & Ni$_2$In (B8$_2$) \\
 CaSn              & tP2  & 123 &     &   &     & 0 & - &     & $\delta$-CuTi (L2$_a$) \\
 \hline
 CuSn$_{4}$        & tP10 & 125 & 550 & - &     &   &   &     & PtPb$_4$ (D1$_d$) \\
 CuSn$_{2}$        & tI12 & 140 &     &   &     & 0 & - &     & Al$_2$Cu (C16) \\
 \hline
 Pd$_{8}$Sn        & aP9  & 2   & 0   & - &     & 0 & - &     & - \\
 Pd$_{6}$Sn        & hR21 & 148 & 0   & - &     & 0 & - &     & - \\
 Pd$_{5}$Sn        & mS12 & 12  & 0   & - & 660 & 0 & - &     & - \\
 Pd$_{5}$Sn$_{3}$  & mS32 & 15  & 0   & - &     &   &   &     & - \\
 PdSn$_{2}$        & tI36 & 140 & 0   & - &     &   &   &     & - \\
 Pd$_{3}$Sn$_{2}$  & oS20 & 36  &     &   &     & 0 & - & 710 & - \\
 PdSn$_{2}$        & tI12 & 140 &     &   &     & 0 & - &     & Al$_2$Cu (C16)\\
 \hline
 AgSn$_{2}$        & tI12 & 140 & 360 & - &     &   &   &     & Al$_2$Cu (C16) \\
 AgSn$_{4}$        & tI10 & 139 & 570 & - &     &   &   &     & Al$_4$Ba (D1$_3$) \\
 Ag$_{7}$Sn        & mP16 & 13  &     &   &     & 0 & - & 880 & - \\
 Ag$_{6}$Sn        & mS28 & 15  &     &   &     & 0 & - &     & - \\
 Ag$_{9}$Sn$_{2}$  & mS44 & 15   &     &   &     & 100 & - &   & - \\
 Ag$_{4}$Sn        & oI40 & 44  &     &   &     & 0 & - &     & - \\
  \hline \hline  
   \end{tabularx} 
  \caption{Compilation of all new thermodynamically stable M-Sn phases predicted in this study. The columns from left to right denote the composition, Pearson symbol, space group number, stable temperature ranges at ambient and elevated pressures, and prototype if available in the AFLOW database.} 
\label{tab:AllNewStrs} 
\end{table}

The systematic scan has uncovered a surprisingly large number of possible thermodynamically stable M-Sn intermetallics summarized in Table~\ref{tab:AllNewStrs}. A total of 11 phases have been shown to be below tie-lines formed by known alloys at ambient pressure and an additional 18 phases have been determined to define convex hulls at 20 GPa. In terms of structural complexity, the identified ground states range from 2- and 3-atom known prototypes with high symmetry (tP2-CaSn and hP3-Na$_2$Sn) to 20- and 22-atom unknown low-symmetry unit cells (mP20-Na$_4$Sn and mS44-Ag$_9$Sn$_2$). According to our experience~\cite{ak23,ak41,ak36,ak44,ak47,ak50}, identification of configurations with over 16 atoms per primitive unit cell is challenging and the conventional global searches performed directly at the DFT level could have easily missed the largest ones located in this study. A common approach for dealing with the problem is to collect relevant known prototypes and carry out a high-throughput chemical substitution screening. We have checked the AFLOW library~\cite{AFLOW, Hicks2021} and found that only 12 out of the 29 predicted structures match the known prototypes in this widely used database. Phases with particular underlying lattices, such as the proposed fcc-based Pd-rich Sn alloys, could also be found with the cluster expansion method but exhaustive screening of possible configurations becomes computationally demanding for large unit cells~\cite{ak47}.

The main findings for each M-Sn binary include the following. In the Na-Sn system, we uncovered three possible high-$T$ ground states at ambient pressure. The identification of the stable hP6-NaSn$_2$ derivative of the hP3-NaSn$_2$ phase is particularly unexpected because the system had been previously explored with global structure searches~\cite{Stratford2017}. The Ca-Sn binary was the only considered system that did not show any new viable ground states at 1 bar. Application of 1.9 GPa pressure induces a transformation of Ca$_2$Sn from oP12 to a more symmetric hP6 polymorph with Ca-Sn honeycomb layers. With the known cP4-CaSn$_3$ remaining stable up to at least 20 GPa, it would be interesting to probe the response of its topological and superconducting properties to compression. In the Cu-Sn binary, we identified a possible high-$T$ ambient-pressure tP10-CuSn$_4$ ground state and observed that the known oS80-Cu$_3$Sn, metastable at ambient conditions~\cite{ak46}, becomes favored at 20 GPa. We also performed a detailed {\it ab initio} analysis of the known alloys and were able to demonstrate how vibrational and configurational entropy terms stabilize off-stoichiometry derivatives of the hP4-CuSn phase at high temperatures. In the Pd-Sn binary, we uncovered several viable ground states across the full composition range, including three fcc-based phases around $x=0.15$ and tI36-PdSn$_2$. To rationalize the stability of the former, we introduced a notation illustrating its morphology in terms of Pd and Sn building blocks and examined the favorability trend of this A$^\circ$B$^\circ$A$^-$A$^\circ$B$^\circ$A$^+$ MSn$_2$ configuration across a set of noble metals. In the Ag-Sn system, we detected several ordered representations of the known disordered alloy in the $0.12<x<0.23$ composition range and identified two possible Sn-rich phases thermodynamically stable at high temperatures. While a number of our previous DFT-based predictions have been validated experimentally, such as the revision of misidentified CrB$_4$~\cite{ak17, ak22}, MnB$_4$~\cite{ak28}, and Na$_2$IrO$_3$~\cite{ak21} crystal structures or the discovery of new LiB~\cite{ak08, ak30}, CaB$_6$~\cite{ak23}, FeB$_4$~\cite{ak16, ak26}, and NaSn$_2$~\cite{ak31,Stratford2017} materials, the feasibility of obtaining the identified thermodynamically stable M-Sn phases may also depend on various kinetic factors that cannot yet be modeled reliably. It will be interesting to see if the appropriate choice of starting materials and/or synthesis conditions could help expand the set of known M-Sn alloys.

We find it important to note that the reliability of these predictions depends strongly on the accuracy of the PBE functional in the GGA chosen for this study. Our additional LDA and SCAN tests presented in Tables S1-S5 illustrate that most of our conclusions are consistent across the considered DFT approximations. Nevertheless, the unexpectedly high number of phases found to be stable in our calculations but never observed experimentally in these extensively studied binaries may still be a consequence of the limited accuracy of standard DFT flavors. Therefore, the dramatically expanded scope of NN-based searches can be instrumental for detecting potential DFT approximation artifacts. Further improvement of NN models may also be needed to ensure a robust identification of all viable candidates. The 8-15 meV/atom accuracy of NN potentials appears to have been appropriate for the considered set of Sn alloys, as our NN+DFT hybrid tests for select Na-Sn and Pd-Sn compositions did not reveal any missed ground states, but more studies should be carried out for materials with diverse bonding types. The encouraging performance of our NN-guided structure prediction approach in the exploration of simple bulk alloys~\cite{ak34,ak37}, metal nanoparticles~\cite{ak38,ak40}, and more complex Sn alloys studied in Ref.~\cite{ak47} and here indicates that it may also be suitable for broader materials classes.

\section*{Data availability statement}
The M-Sn NN models, IDs 002BD87A, 0552A1B2, 01121374, 036C3A42, 016C1E4A, can be downloaded at \url{https://github.com/maise-guide/maise/}. Relevant M-Sn structures are given in the supplementary material. Other data supporting the ﬁndings of this study is available from the corresponding
author upon request.

\section*{Code availability statement}
The {\small MAISE} and {\small MAISE-NET} codes are freely available for download at \url{https://github.com/maise-guide/}.

\section*{Acknowledgements} 
We acknowledge the NSF support (Award No. DMR-1821815) and the Extreme Science
and Engineering Discovery Environment computational resources~\cite{exsede} (NSF Award
No. ACI-1548562, Project No. TG- PHY190024).  

\section*{Competing interests}
The authors declare no competing interests.

\section*{Additional information}
\textbf{Supplementary information} The online version contains supplementary 
material. 

\bibliographystyle{naturemag}
\bibliography{main}

\end{document}